\documentclass[useAMS,onecolumn]{mn2e}
\usepackage{graphics,graphicx}
\usepackage{color}
\usepackage{epsf}
\usepackage{epsfig}
\usepackage{dcolumn}

\newcommand{\mdot}{\mbox{$\dot{M}$}}

\newcommand{\lsim}{\raisebox{-.4ex}{$\stackrel{<}{\scriptstyle \sim}$}}
\newcommand{\gsim}{\raisebox{-.4ex}{$\stackrel{>}{\scriptstyle \sim}$}}

\usepackage{txfonts}
\title[]{Two temperature accretion around rotating black holes: Description of
general advective flow paradigm in presence of various cooling processes to
explain low to high luminous sources}

\author[]{\Large S. R. Rajesh\thanks{rajesh@physics.iisc.ernet.in}, Banibrata 
Mukhopadhyay\thanks{bm@physics.iisc.ernet.in}\\
Astronomy and Astrophysics Program, Department of Physics,
Indian Institute of Science, Bangalore 560012, India\\ 
}

\begin{document}


\maketitle
\label{firstpage}


\begin{abstract}
We investigate the viscous two temperature accretion disc flows
around rotating black holes. We describe the global solution 
of accretion flows with a sub-Keplerian angular momentum profile, 
by solving the underlying conservation equations 
including explicit cooling processes selfconsistently. 
Bremsstrahlung, synchrotron and inverse Comptonization of soft photons 
are considered as possible cooling mechanisms. We 
focus on the set of solutions for sub-Eddington, Eddington and super-Eddington 
mass accretion rates around Schwarzschild and Kerr black holes with a Kerr
parameter $0.998$. It is found that the flow, during its infall from 
the Keplerian to sub-Keplerian transition region to the black hole
event horizon, passes through various phases of advection -- general advective  
paradigm to radiatively inefficient phase and vice versa. Hence the flow governs much
lower electron temperature $\sim 10^8-10^{9.5}$K, 
in the range of accretion rate in Eddington units 
$0.01\lsim\mdot\lsim 100$, compared to the hot protons
of temperature $\sim 10^{10.2}-10^{11.8}$K. Therefore, the solution may potentially
explain the hard X-rays and $\gamma$-rays emitted from AGNs and X-ray binaries.
We then compare the solutions for two different regimes of viscosity and
conclude that a weakly viscous flow is expected to be cooling dominated,
particularly at the inner region of the disc, compared to its highly viscous
counter part which is radiatively inefficient. With all the solutions in hand,
we finally reproduce the observed luminosities of the under-fed AGNs and
quasars (e.g. Sgr~$A^*$) to ultra-luminous X-ray sources (e.g. SS433), 
at different combinations of input parameters such as mass accretion rate, ratio of
specific heats.
The set of solutions also predicts appropriately the luminosity 
observed in the highly luminous AGNs and ultra-luminous quasars (e.g. PKS~0743-67).

\end{abstract}


\section{Introduction}

The cool Keplerian accretion disc (Pringle \& Rees 1972; 
Shakura \& Sunyaev 1973; Novikov \& Thorne 1973)
was found to be inappropriate to explain observed hard X-rays, 
e.g. from Cyg~X-1 (Lightman \& Shapiro 1975). It was argued
that secular instability of the cool disc swells the optically
thick, radiation dominated region to a hot, optically thin, gas 
dominated region resulting in hard component of spectrum $\sim 100$KeV
(Thorne \& Price 1975; Shapiro, Lightman \& Eardley 1976). This
region is strictly of two temperatures with electron and ion
temperatures respectively $\sim 10^9$K and $\sim 5\times 10^{11}$K
which confirms that cool, one temperature, pure Keplerian accretion 
solution is not unique. Indeed Eardley \& Lightman (1975) found 
that a Keplerian disc is unstable due to thermal and viscous effects
when viscosity parameter $\alpha$ (Shakura \& Sunyaev 1973) 
is constant. Later Eggum et al. (1985)
showed by numerical simulations that the Keplerian disc with a
constant $\alpha$ collapses. 

Around eighties, therefore, the idea of two component accretion disc
started floating around. For example, Paczy\'nski \& Wiita (1980)
described a geometrically thick regime of the accretion disc in the optically thick
limit, while Rees et al. (1982) introduced accretion 
torus in the optically thin limit. Moreover the idea of sub-Keplerian, 
transonic accretion was introduced by Muchotrzeb \& Paczy\'nski (1982),
which was later improved by other authors (Chakrabarti 1989, 1996; 
Mukhopadhyay 2003). Other models were proposed by e.g. Gierli\'nski et al. (1999), 
Coppi (1999), Zdziarski et al. (2001), including a secondary component in
the accretion disc. On the other hand, Narayan \& Yi (1995) 
introduced a two temperature disc model in the regime of inefficient cooling
resulting in a vertical thickening of the hot disc gas. Here the pressure
forces are expected to become important in modifying the disc dynamics
which is likely to be sub-Keplerian. Other models with similar properties
were proposed by, e.g., Begelman (1978), Liang \& Thompson (1980), Rees et al. (1982),
Eggum, Coroniti \& Katz (1988). Abramowicz et al. (1988) proposed a height-integrated
disc model, namely ``slim disc", having high optical depth of the accreting gas
at super-Eddington accretion rate such that the diffusion time is longer than
the viscous time. The model was further applied to study the thermal and viscous 
instabilities in optically thick accretion discs (Wallinder 1991; Chen \& Taam 1993).

Shapiro, Lightman \& Eardley (1976) initiated a two temperature Keplerian 
accretion disc at a low mass accretion rate which is optically thin and 
significantly hotter than the single temperature Keplerian disc of Shakura \& Sunyaev (1973).
The optically thin hot gas cools down through the bremsstrahlung and inverse-Compton
processes and could explain various states of Cyg~X-1 (Melia \& Misra 1993).
Similarly, the ``ion torus" model by Rees et al. (1982) was applied to explain 
AGNs at a low mass accretion rate. However, the two temperature model solutions by
Shapiro, Lightman \& Eardley (1976) appear thermally unstable. 
Narayan \& Popham (1993) and subsequently Narayan \& Yi (1995) showed that
introduction of advection may stabilize the system. However, the solutions
of Narayan \& Yi (1995), while of two temperatures, 
could explain only a particular class of hot systems with inefficient 
cooling mechanisms. They also described the hot flow based on the assumption of
``self similarity" which is just a ``plausible choice".
They kept the electron heating decoupled 
from the disc hydrodynamical computations which merely is an assumption. Later on,
the solutions were attempted to generalize by Nakamura et al. (1997),
Manmoto et al. (1997), Medvedev \& Narayan (2001),
relaxing efficiency of cooling into
the systems, but concentrating only on specific classes of solutions.
On the other hand, Chakrabarti \& Titarchuk (1995) and
later Mandal \& Chakrabarti (2005) modeled two temperature accretion
flows around Schwarzschild black holes in the general ``advective paradigm",
emphasizing possible formation of shock and its consequences
therein. However, they also did not include the effect of electron heating 
self-consistently into the hydrodynamical equation, and thus 
the hydrodynamical
quantities do not get coupled to the rate of electron heating (see also Rajesh 
\& Mukhopadhyay 2009). 

In the present paper, we model a selfconsistent accretion flow in 
the regime of two temperature transonic sub-Keplerian disc (see also
Sinha, Rajesh \& Mukhopadhyay 2009; a brief version of the
present work, but around Schwarzschild black holes). We consider all
the hydrodynamical equations of the disc along with thermal components
and solve the coupled set of equations selfconsistently. We neither
restrict to the advection dominated regime nor the self-similar solutions.
We allow the disc to cool selfconsistently according to the 
thermo-hydrodynamical evolution and compute the corresponding
cooling efficiency factor as a function of radial coordinate. We investigate that
when does the disc switch from the radiatively inefficient nature to general
advective paradigm and vice versa.

In order to implement our model to explain observed sources,
we focus on the under-luminous AGNs and quasars (e.g. Sgr~$A^{*}$),
ultra-luminous quasars and highly luminous AGNs 
(e.g. PKS~0743-67) and
ultra-luminous X-ray (ULX) sources (e.g. SS433), when the last items  
are likely to be the ``radiation trapped'' accretion discs. 
While the first two cases correspond to respectively 
sub-Eddington and super-Eddington accretion flows around
supermassive black holes, the last case corresponds to super-Eddington 
accretors around stellar mass black holes. 

In the next section, we discuss the model equations describing the
system and the procedure to solve them. Subsequently, we discuss
the two temperature accretion disc flows around stellar mass
and supermassive black holes, respectively in \S3 and \S4,
for both sub-Eddington, Eddington and super-Eddington accretion rates.
Section 5 compares the disc flow of low Shakura-Sunyaev (1973) $\alpha$ with 
that of high $\alpha$ and then between the flows around co
and counter rotating black holes. Then we discuss the implications of the results
with a summary in \S6.

\section{Model equations describing the system and solution procedure}

For the present purpose, we set five coupled differential equations
describing the law of conservation in the sub-Keplerian optically thin accretion regime. 
Necessarily the set of equations describes the inner part of the accretion
disc where the gravitational potential energy dominates over the 
centrifugal energy of the flow. 

Throughout, we express all the
variables in dimensionless units, unless stated otherwise.
The radial velocity $\vartheta$ and sound speed $c_s$ are expressed 
in units of light speed $c$, the specific angular momentum $\lambda$
in $GM/c$, where $G$ is the Newton's gravitational constant and $M$ is the
mass of the compact object, for the present purpose black hole, 
expressed in units of solar mass $M_\odot$, the 
radial coordinate $x$ in units of $GM/c^2$, the density $\rho$ and the total pressure $P$
accordingly. The disc fluid under consideration consists of ions and
electrons --- thus two fluid/temperature system, apart from radiation. 
Furthermore, at the high temperature, the disc flow with ions/electrons
behaves as (almost) noninteracting gas.

\subsection{Conservation laws }
(a) Mass transfer:
\begin{eqnarray}
\frac{1}{x} \frac{\partial}{\partial x} (x \rho \vartheta)  =  0,
\label{eoc}
\end{eqnarray}
whose integrated form gives the mass accretion rate
\begin{eqnarray}
\mdot\,=\,-4\pi x \Sigma \vartheta,
\label{mass}
\end{eqnarray}
where the surface density
\begin{eqnarray}
\Sigma \,= \,I_n\, \rho h(x), 
\label{sden}
\end{eqnarray}
\begin{eqnarray}
I_n \,= \,(2^{n} n!)^{2}/(2n+1)!\,\, {\rm (Matsumoto\hskip0.2cm et\hskip0.2cm al.\hskip0.2cm 1984)},
\label{matsu}
\end{eqnarray}
$n$ is the polytropic index which is equal to $1/(\gamma-1)$ when $\gamma$ is the ratio of specific
heats, and half-thickness, based on the vertical equilibrium assumption, of the disc
\begin{eqnarray}
h(x) \,= \,c_s x^{1/2} F^{-1/2}.
\label{thik}
\end{eqnarray}

(b) Radial momentum balance:
\begin{eqnarray}
\vartheta \frac{d\vartheta}{dx} \,+ \,\frac{1}{\rho} \frac{dP}{dx} \,- \,\frac{\lambda^{2}}{x^3} \,+ \,F \, = \, 0
\label{rad}
\end{eqnarray}
when following pseudo-Newtonian approach of Mukhopadhyay (2002)
\begin{eqnarray}
F=\frac{(x^2 - 2a\sqrt{x} +a^2)^2}{x^3[\sqrt{x}(x - 2) +a]^2},
\end{eqnarray}
where $a$ is the specific angular momentum (Kerr parameter) of the black hole.
We also define a parameter 
\begin{eqnarray}
\beta=\frac{{\rm gas\hskip0.2cm pressure\hskip0.2cm} P_{gas}}
{{\rm total\hskip0.2cm pressure\hskip0.2cm} P}\,=\frac{6\gamma-8}{3(\gamma-1)}
\,\,\,\, {\rm (e.g.\hskip0.2cm Ghosh\hskip0.2cm \&\hskip0.2cm Mukhopadhyay\hskip0.2cm 2009)},
\end{eqnarray}
where $\gamma$ may range from
$4/3$ to $5/3$, $P_{gas}=P_i\hskip0.2cm({\rm ion\hskip0.2cm pressure})
+P_e\hskip0.2cm({\rm electron\hskip0.2cm pressure})$, such that
\begin{eqnarray}
P \,= \,\frac{\rho}{\beta\,c^2} \left(\frac{kT_{i}}{\mu_{i} m_{i}} \,+ 
\,\frac{kT_{e}}{\mu_{e} m_{i}}\right) \,= \,\rho\, c^{2}_{s},
\label{ptot}
\end{eqnarray}
where $T_{i}$, $T_{e}$ are respectively the ion and electron temperatures in Kelvin, 
$m_i$ is the mass of proton in gm, $\mu_i$ and $\mu_e$
respectively are the corresponding effective molecular weight, $k$ the Boltzmann constant.  
We assume $\beta$ (and then $\gamma$) constant throughout the flow.

(c) Azimuthal momentum balance:

\begin{eqnarray}
\vartheta \frac{d \lambda}{dx} \,= \,\frac{1}{\Sigma x} \frac{d}{dx}\left(x^2 |W_{x \phi}|
\right), 
\label{az}
\end{eqnarray}
where following Mukhopadhyay \& Ghosh (2003; hereinafter MG03) the shearing stress can be
expressed in terms of the pressure and density as
\begin{eqnarray}
W_{x \phi}  \,= \,- \alpha  \left(I_{n+1} P_{eq} \,+ \,I_n \vartheta^2 \rho_{eq} \right)h(x), 
\end{eqnarray}
where $\alpha$ is the dimensionless viscosity parameter and $P_{eq}$ and $\rho_{eq}$ are
the pressure and density respectively 
at the equatorial plane. Note that we will assume $P_{eq}\sim P$ and
$\rho_{eq}\sim\rho$ in obtaining solutions.

(d) Energy production rate:

\begin{eqnarray}
\frac{\vartheta h(x)}{\Gamma_{3} - 1} \left(\frac{dP}{dx} \,- \,\Gamma_{1} \frac{P}{\rho} \frac{d \rho}{dx}\right) \,= \,Q^{+} \,- \,Q_{ie},
\label{eni}
\end{eqnarray}
where following MG03
\begin{eqnarray}
Q^+ \,= \, \alpha (I_{n+1} P \,+ \,I_n \vartheta^2 \rho )h(x) \frac {d \lambda}{dx},
\label{qvis}
\end{eqnarray}
which is the heat generated by viscous dissipation, and $Q_{ie}$ is
the Coulomb coupling (Bisnovatyi-Kogan \& Lovelace 2000) given in dimensionful
unit as
\begin{eqnarray}
\nonumber
&&q_{ie} \, = \,\frac{8 (2 \pi)^{1/2}e^4 n_i n_e}{m_i m_e}\left(\frac{T_e}{m_e} \,+ \, \frac{T_i}{m_i}\right)^{-3/2} \ln (\Lambda) \ \left(T_i \, - \,T_e \right)\,\,{\rm erg/cm^3/sec}s\\
&&{\rm when}\,\,\,q_{ie}=Q_{ie}\,c^{11}/(h\,G^4\,M^3).
\label{qie}
\end{eqnarray}
Here $n_i$ and $n_e$ denote number densities of ion and electron respectively, 
$e$ the charge of an electron,
ln($\Lambda$) the Coulomb logarithm.
We also define (MG03)
\begin{eqnarray}
\Gamma_{3} &= &1 \,+ \,\frac{\Gamma_{1} \,- \,\beta}{4 \,- \,3 \beta},\\
\Gamma_{1} &= &\,\beta \,+ \,\frac{(4 \,- \,3 \beta)^{2}(\gamma \,- \,1)}{\beta+\,12(\gamma \,- \,1)(1 \,- \,\beta)}.
\label{gam}
\end{eqnarray}

(e) Energy radiation rate:

\begin{eqnarray}
\frac{\vartheta h(x)}{\Gamma_{3} - 1} \left(\frac{dP_{e}}{dx} \,- \,\Gamma_{1} \frac{P_{e}}{\rho} \frac{d \rho}{dx}\right) \,= \,Q_{ie} \,-\,Q^{-},
\label{ene}
\end{eqnarray}
where $Q^-$ is the heat radiated away by the
bremsstrahlung ($q_{br}$), synchrotron ($q_{syn}$) processes and inverse Comptonization
($q_{comp}$) due to soft synchrotron photons, given in dimensionful form as
\begin{eqnarray}
q^-=q_{br}+q_{syn}+q_{comp},\,\,\,{\rm when}\,\,\,q^-=Q^-\,c^{11}/(h\,G^4\,M^3).
\label{qm}
\end{eqnarray}
Various components of the cooling processes may be read as (see Narayan \& Yi 1995; 
Mandal \& Chakrabarti 2005 for detailed description, what we do not repeat here)
\begin{eqnarray}
\nonumber
q_{br} &= &1.4 \times 10^{-27} \ n_e\,n_i T_e^{1/2}\,(1+4.4\times 10^{-10} T_e)\,
\,\,{\rm erg/cm^3/sec},\\
\nonumber
q_{syn} & = &\frac{2 \pi}{3 c^2} kT_e \, \frac{\nu_a^{3}}{R}\,\,\,{\rm erg/cm^3/sec},
\,\,\,R=x\,GM/c^2,\\
\nonumber
q_{comp} &=& {\cal F}\,q_{syn},\,\,\ {\cal F} = \eta _{1}\,
\left(1 \,- \,\left(\frac{x_{a}}{3 \theta _{e}}\right)^{\eta _{2}} \right),\,\,\,
\eta_{1} = \frac{p(A-1)}{(1-pA)},\,\,\
p \,= \,1 \,- \,\exp(- \tau _{es}),\\
A &= &1 \,+ \,4 \theta_{e} + \,16 \theta^{2}_{e},\,\,\,\theta_{e} \,= 
\,kT_{e}/m_{e} c^{2},\,\,\
\eta_{2} =  -\left(1 \,+ \,\frac{ln(p)}{ln(A)}\right),\,\,\,x_{a} \,= \,h \nu_{a}/m_{e} c^{2},
\label{qvari}
\end{eqnarray}
where $\tau_{es}$ is the scattering optical depth given by
\begin{eqnarray}
\tau_{es}=\kappa_{es}\rho\,h
\end{eqnarray}
where $\kappa_{es}=0.38$ cm$^2$/gm and $\nu_a$ is the 
synchrotron self-absorption cut off frequency determined by following Narayan \& Yi (1995).
Note that without a satisfactory knowledge of the magnetic field in accretion disks,
following Mandal \& Chakrabarti (1995), we assume the maximum possible magnetic energy
density to be the gravitational energy density of the flow.
As the total optical
depth should include the effects of absorption due to nonthermal processes,
effective optical depth is computed as
\begin{eqnarray}
\tau_{\rm eff}\simeq\sqrt{\tau_{es}\,\tau_{abs}}
\end{eqnarray}
where 
\begin{eqnarray}
\tau_{abs}=\frac{h}{4\sigma T_e^4}\left(q_{br}+q_{syn}+q_{comp}\right)\frac{GM}{c^2},
\end{eqnarray}
when $\sigma$ is Stefan-Boltzmann constant.

Now combining all the above equations we obtain
\begin{eqnarray}
\frac{d\vartheta}{dx} \,= \,\frac{N(x,\vartheta,c_{s},\lambda,T_{e})}{D(\vartheta,c_{s})},
\label{dvdx}
\end{eqnarray}
where
\begin{eqnarray}
\nonumber
N \,= \,\frac{\Gamma_1 \,+ \,1}{\Gamma_3 \,- \,1} \vartheta^{2} c_s J\,- \,
\frac{ \alpha^{2}  c_s}{x} H\left( \frac{I_{n+1}}{I_n} c_s^{2} \,+ \,\vartheta^{2} \right) \,- \,
\alpha^{2} \frac{I_{n+1}}{I_n} 2H J\,+ \,
  \frac{\Gamma_1 \,- \,1}{\Gamma_3 \,- \,1} \vartheta^{2} c_s^{3} L \,
+ \,\alpha H \left(\frac{2\lambda \vartheta c_s}{x^{2}} \right) \\
+ \,\frac{4\pi Q_{ie}}{\mdot}\vartheta^{2} c_s^{2} x^{3/2} F^{-1/2},
\label{num}
\end{eqnarray} 
\begin{eqnarray}
D \,= \,\frac{1 \,- \,\Gamma_1 }{\Gamma_3 \,- \,1} c_s^{3} \vartheta \,+ \,2 \alpha c_s \frac{I_{n+1}}{I_n}H \left(\frac{c_s^{2}}{\vartheta} \,- \,\vartheta \right) \,+ \,\frac{\Gamma_1 \,+ \,1}{\Gamma_3 \,- \,1} \vartheta^{2} c_s^{2} \left(\vartheta \,- \,\frac{c_s^{2}}{\vartheta}\right) \,+ \,\alpha^{2} \vartheta H \left(\frac{H}{\vartheta}\right)
\label{den}
\end{eqnarray}
and
\begin{eqnarray}
L \,= \,\left(\frac{3}{2x} \,- \,\frac{1}{2F}\frac{dF}{dx} \right),\,\,
H \,= \,\left( I_{n+1} c_s^{2} \,+ \,I_n \vartheta^{2} \right),\,\,
J \,= \,\left( c_s^{2} L \,+ \,\frac{\lambda^{2}}{x^{3}} \,- \,F\right).
\end{eqnarray}

We know that around the sonic radius $N=D=0$ (Mukhopadhyay 2003) and hence obtain the Mach number
at this critical radius from $D=0$
\begin{eqnarray}
M_c \,= \frac{\vartheta_c}{c_{sc}}\,=
\,\sqrt{\frac{-B + \left(B^{2} \,- \,4AC \right)^{1/2}}{2A}},
\end{eqnarray}
where
\begin{eqnarray}
\nonumber
A &= &\frac{\Gamma_1 \,+ \,1}{1 \,- \,\Gamma_3} \,- \,2 \alpha^{2} \left(I_{n+1} \,- \,I_{n} \right),\,\,
B \,= \,\frac{2 \left(\Gamma_1 \,+ \,1 \right)}{1 \,- \,\Gamma_3} \,- \,2 \alpha^{2} \frac{I_{n+1}}{I_n} \left(I_{n+1} \,- \,I_{n} \right),\\
C &= &\alpha^{2} \frac{I_{n+1}}{I_n} \left(1 \,- \,2 I_{n+1}\right).
\end{eqnarray}
Also from $N=0$, we can compute explicitly $c_{sc}$ as a function
of sonic/critical radius $x_c$. For a physical $x_c$, what one has
to adjust in order to obtain a physical solution connecting outer
boundary to black hole horizon through $x_c$, $c_{sc}$ and then 
$\vartheta_c$ can be assigned, discussed in APPENDIX A in detail.
Note that an improper $x_c$ may lead to an unphysical/imaginary
$c_{sc}$ and $\vartheta_c$.

Finally combining Eqns. (\ref{rad}), (\ref{az}) and (\ref{ene}) we obtain
\begin{eqnarray}
\frac{d c_{s}}{dx} \,= \,\left(\frac{c_{s}}{\vartheta} - \frac{\vartheta}{c_{s}}\right) \frac{d\vartheta}{dx} \,+
\,\frac{J}{c_{s}},
\end{eqnarray}
\begin{eqnarray}
\frac{d \lambda}{dx} \,= \,\left(\frac{2 \alpha x}{\vartheta c_{s}} \frac{I_{n+1}}{I_{n}}
\left(\frac{c_{s}^{3}}{\vartheta}-\vartheta c_{s}\right)+\alpha x\right) \frac{d\vartheta}{dx} \,+
\,\left(\frac{c_{s}^{2}-2x \alpha J}{c_{s}}+\vartheta\right),
\end{eqnarray}
\begin{eqnarray}
\frac{dT_{e}}{dx} \,= \,(1- \Gamma _{1})T_{e} \frac{\vartheta}{c_{s}^{2}} \frac{d\vartheta}{dx} \,+ \,(1-
\Gamma _{1})T_{e}\left(\frac{J}{c_{s}^{2}}+L\right) \,+ \,\frac{(\Gamma_{3}-1)4
\pi}{\mdot}\frac{c_{s} x^{3/2}}{F^{1/2}} \left(Q^{ie}-Q^{-}\right). 
\end{eqnarray}
Hence knowing $\vartheta$ we can obtain other variables $c_s$, $\lambda$, $T_e$.
Note that $d\vartheta/dx$ is indeterminate (of $0/0$ form) at $x_c$.
APPENDIX B discusses the procedure to obtain $d\vartheta/dx$ at $x_c$.

As the entropy increases inwards in advective flows (e.g. Narayan \& Yi 1994;
Chakrabarti 1996; MG03), there is a possibility of convective instability 
and then corresponding transport, as proposed by Narayan \& Yi (1994).
Dynamical convective instability arises when the square of effective frequency 
\begin{eqnarray}
\nu_{\rm eff}^{2} \,= \,\nu_{\rm BV}^{2} \,+ \,\nu_r^{2}\,<\,0
\end{eqnarray}
such that $\nu_{\rm BV}$ is the Brunt-V\"ais\"al\"a frequency given by
\begin{eqnarray}
\nu_{\rm BV}^{2} \,= \,-\frac{1}{\rho} \frac{dP}{dx} \frac{d}{dx}ln\left(\frac{P^{1/
\gamma}}{\rho}\right)
\end{eqnarray}
and $\nu_r$ is the radial epicyclic frequency.

\subsection{Solution procedure}
\begin{figure}
\centering
\includegraphics[width=0.70\columnwidth]{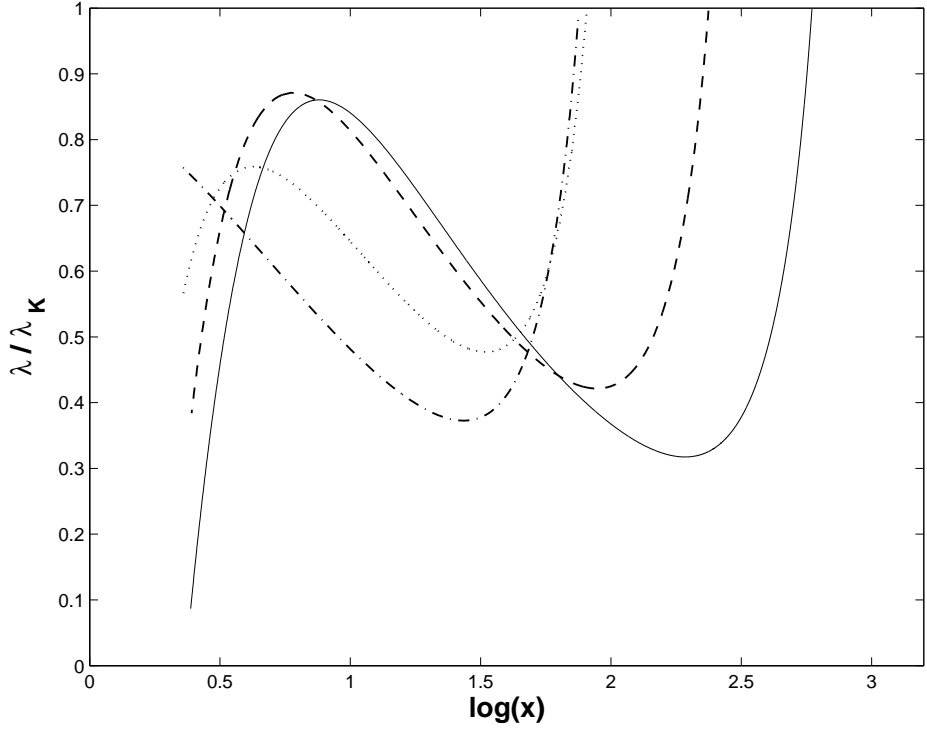}
\caption{
Variation of ratio of disc specific angular momentum to corresponding 
Keplerian angular momentum as a function of radial coordinate, when
solid, dashed, dotted, dot-dashed lines correspond to the cases
with $a=-0.5,0,0.5,0.998$ respectively.
Other parameters are $\mdot=1$, $M=10$, $\alpha = 0.01$.
 }
\label{figlam} \end{figure}

In order to obtain the steady state solution, as of previous work (MG03, Mukhopadhyay 2003), 
primarily we need
to fix the appropriate critical radius $x_c$ (in fact the energy at the critical radius
which is not conserved in the present cases) and 
the corresponding specific angular momentum $\lambda_c$ of the flow. 
The detailed description of the procedure to obtain physically meaningful values of 
$x_c$ and $\lambda_c$, to be determined iteratively, 
is given in APPENDIX A. As the flow is considered to be
of two temperatures, at $x_c$ an appropriate electron temperature $T_{ec}$ also needs to be
determined; also discussed in APPENDIX A. 
Note that one has to adjust the set of values $x_c, \lambda_c, T_{ec}$ appropriately/iteratively
to obtain self-consistent solution connecting
outer boundary and black hole event horizon through $x_c$. 
Depending on the type of accreting 
system to model, we then have to specify the related
inputs: $\mdot$, $M$, $\gamma$ and $a$. 
Important point to note is that unlike former works (e.g. Chakrabarti \& Titarchuk 1995,
Chakrabarti 1996, MG03) here $x_c$ changes with the change of $\mdot$,
because the various cooling processes considered here explicitly depend on $\mdot$.
Finally, we have 
to solve the Eqn. (\ref{dvdx}) from $x_c$ to inwards --- upto the 
black hole event horizon, and to outwards --- upto the transition 
radius $x_o$ where the disc
deviates from the Keplerian to the sub-Keplerian regime such that $\lambda/\lambda_K=1$
($\lambda_K$ being the specific angular momentum of the
Keplerian part of the disc). 
Figure \ref{figlam} shows how the ratio $\lambda/\lambda_K$ varies as a function
of radial coordinate for different $a$. Note that higher $a$, which
corresponds to a lower disc angular momentum (Mukhopadhyay 2003), reassembles 
the Keplerian part advancing with a smaller size of the sub-Keplerian disc. 
On the other hand, for a lower $a$ the inner edge of the Keplerian component
recedes. The fact of moving in and out of the inner edge of the disc 
reassembles respectively the soft and hard state of the black hole 
(e.g. Gilfanov et al. 1997). Hence it is naturally
expected to link with the spin of the black hole. 

However, the important point
to note is that there is no selfconsistent model to describe the transition region 
where $\lambda/\lambda_K=1$. Therefore, the transition of the flow from the Keplerian to
sub-Keplerian regime does not appear smooth. This is mainly because the set of
equations used to model the sub-Keplerian flow is not valid to 
explain the cold Keplerian flow, unless an extra boundary condition is imposed 
at the outer edge of the sub-Keplerian disc. However, in the present paper
we do not intend to address the transition zone; rather 
we prefer to stick with the sub-Keplerian flow.
Narayan et al. (1997) imposed 
boundary conditions at both the ends of accretion flows along with at the
critical radius to fix the problem, at the cost of more input parameters than
the parameters chosen in the present work. But still the transition of the 
flow from the Keplerian to sub-Keplerian regime remains undefined.
Later Yuan (1999) discussed
how the solutions vary with the change of outer boundary conditions influencing
the structure of an optically thin accretion flow.

Below we discuss solutions in various parameter regimes to understand properties
of the accretion disc around, first, stellar mass ($M=10$) and then
super-massive ($M=10^7$) black holes.

\section{Two temperature accretion disc around stellar mass black holes}

Primarily we concentrate on two extreme regimes: (1) sub-Eddington and 
Eddington limits of accretion, (2) super-Eddington accretion. 
Furthermore, at each case of accretion rate, we focus on
solutions around nonrotating (Schwarzschild) and rotating 
(Kerr with $a=0.998$) black holes. 

One of our aims is to understand how the explicit cooling processes affect
the disc dynamics and then the cooling efficiency vary over the disc radii. 
The cooling efficiency $f$ is defined as the ratio of the energy advected by the
flow to the energy dissipated, which is $1$ for
the advection dominated accretion flow (in short ADAF; 
Narayan \& Yi 1994, 1995) and less than
$1$ for the general advective accretion flow (in short GAAF; Chakrabarti 1996; 
Mukhopadhyay 2003; MG03) in general.
Therefore, $f$ directly controls the ion and electron temperatures
of the disc. 
Far away from the black hole where the gravitational power
is weaker, the angular momentum profile becomes Keplerian 
and thus the disc becomes (or tends to become) of one
temperature in the presence of efficient cooling. 

\subsection{Sub-Eddington and Eddington accretors}

\subsubsection{Schwarzschild black holes}

\begin{figure}
\centering
\includegraphics[width=0.70\columnwidth]{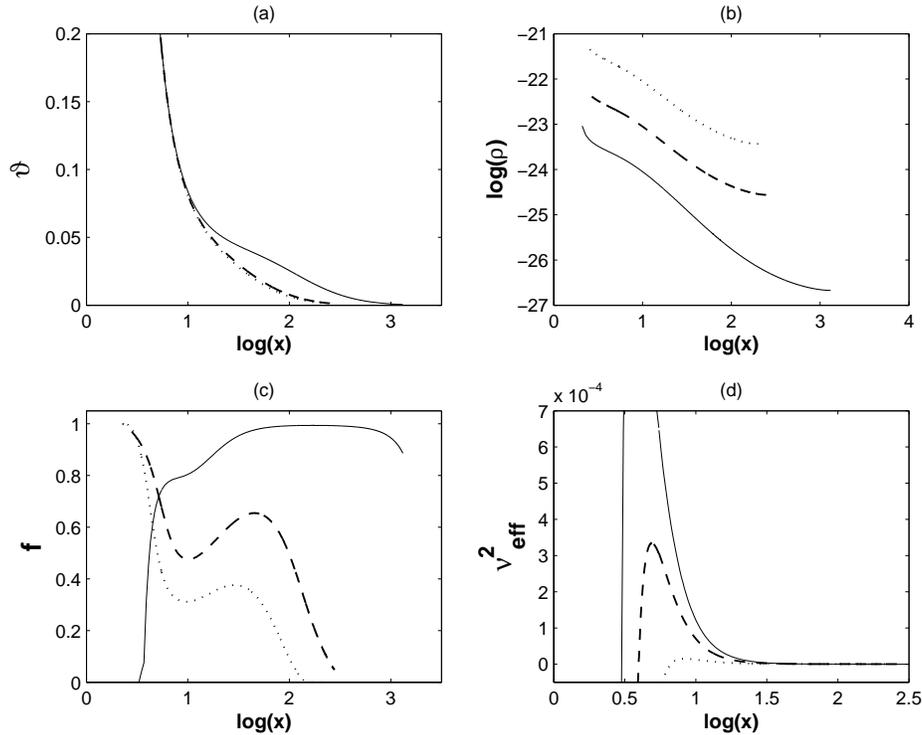}
\caption{
Variation of dimensionless (a) radial velocity, (b) density,
(c) cooling factor,
(d) square of convective frequency, as functions of radial 
coordinate for sub-Eddington and Eddington accretion flows.
Solid, dashed, dotted curves are for $\mdot=0.01,0.1,1$ respectively.
Other parameters are $a=0$, $\alpha = 0.01$, $M=10$; see Table 1 for details.
 }
\label{figstl0} \end{figure}

We first consider flows around static black holes where the Kerr parameter $a=0$.
Figure \ref{figstl0} shows the behavior of flow variables as functions of
radial coordinate for $\mdot=0.01,0.1,1$; throughout in the text we
express $\mdot$ in units of Eddington limit. The sets of input parameters
for the model cases described here are given in Table 1. Figure \ref{figstl0}a
\newpage
\noindent{ Table 1: Parameters for accretion with $\alpha=0.01$ around 
black holes of $M=10$, when the subscript `c' indicates the quantity at the
critical radius and $T_{ec}$ is expressed in units of $m_i c^2/k$ }

\begin{center}
\begin{tabular}{lllllllllllll}
\hline
\hline
$\mdot$ & $a$ & $\gamma$ & $x_{c}$ & $\lambda _{c}$ & $T_{ec}$  \\
\hline
\hline
& & Sub-Eddington, & Eddington & accretors & \\
\hline
\hline
0.01 & 0     & 1.5  & 5.5 & 3.2 & 0.0001 \\
0.01 & 0.998 & 1.5  & 3.5 & 1.7 & 0.0001 \\
\hline
0.1  & 0     & 1.4  & 5.5 & 3.2 & 0.000164\\
0.1  & 0.998 & 1.4  & 3.5 & 1.7 & 0.000153\\
\hline
1    & 0     & 1.35 & 5.5 & 3.2 & 0.000225\\
1    & 0.998 & 1.35 & 3.5 & 1.7 & 0.0002122\\
\hline
\hline
& & Super-Eddington & accretors & & \\
\hline
\hline
10   & 0     & 1.345& 5.5 & 3.2 & 0.000181565\\
10   & 0.998 & 1.345& 3.5 & 1.7 & 0.0004432\\
\hline
100  & 0     & 1.34 & 5.5 & 3.2 & 0.00038678\\
100  & 0.998 & 1.34 & 3.5 & 1.7 & 0.00055\\
\hline
\hline

\end{tabular}
\end{center}

verifies that a higher radial velocity corresponds to a lower mass accretion rate
of the flow ($\sim 0.01$) which results in less possible accumulation of matter in 
a particular radius attributing to a lower disc density (Fig. \ref{figstl0}b). 
This hinders the bremsstralung process to cool the flow.
However, at around
$x=30$ the centrifugal barrier dominates and brings the velocity $\vartheta$ down,
particularly for $\mdot=0.01$,
which finally merges with that of higher $\mdot$-s. On the other hand,
a lower $\mdot$ corresponds to a gas dominated hot flow, which is 
radiatively less efficient and quasi-spherical in nature. As a result
$\vartheta$ is high, as seen in Fig. \ref{figstl0}a. Efficiency of cooling 
is shown in Fig. \ref{figstl0}c. Naturally a low $\mdot$ corresponds
to a radiatively inefficient flow rendering $f\gsim 0.9$ upto $x=30$.
At $x<30$, the dominance of centrifugal barrier slows down the infall
which increases the residence time of matter in the disc before plunging into
the black hole. This allows matter to have enough time to radiate 
by the synchrotron process and inverse Comptonization due to
synchrotron soft photons, rendering $f\rightarrow 0$ close to the black hole. 
In other words, for $\mdot=0.01$, the disc is essentially radiatively 
inefficient, upto $x\sim 30$, and therefore the electron temperature never goes down. However,
the density sharply increases in the vicinity of the black hole (Fig. \ref{figstl0}b)
which favors efficient cooling at a high temperature.


Therefore, although far away from the black hole a sub-Eddington flow 
appears to be radiatively inefficient, from $x=30$ onwards it turns
out to be a radiatively efficient advective flow with $f$ much less than unity. 
However, for $\mdot=0.1,1$ the density is higher than that for $\mdot=0.01$
and hence the bremsstrahlung effect starts playing role in radiation mechanisms
at much outer radii.
This decreases the ion-electron temperature difference at the transition radius $x_o$.
However, as the flow advances the synchrotron and corresponding inverse Compton
effects dominate attributing to strong radiative
loss. This renders $f\lsim 0.5$ upto $x=10$. Further in, a strong
radial infall, in absence of any centrifugal barrier compared to a low $\mdot$
case, does not permit matter to radiate enough, 
rendering $f$ upto $1$. This is particularly because the strong advection decreases
the residence time of the flow before plunging into the black hole and thus
renders a weaker ion-electron coupling. This in turn hinders the transfer 
of energy from the ions to electrons attributing ions to remain hot, while
electrons continue to be cooled down further by radiative processes.

However, Fig. \ref{figstl0}d shows that either of the cases do not exhibit convective 
instability (see, however, Narayan, Igumenshchev \& Abramowicz 2000,
Quataert \& Gruzinov 2000) upto very inner edge, 
evenif the radiatively inefficient flow deviates to a radiatively efficient
GAAF. At a very inner edge, discs with
$\mdot=0.1,1$ particularly appear to be marginally unstable, which, although,
seems not playing any role in angular momentum transfer.

\begin{figure}
\centering
\includegraphics[width=0.80\columnwidth]{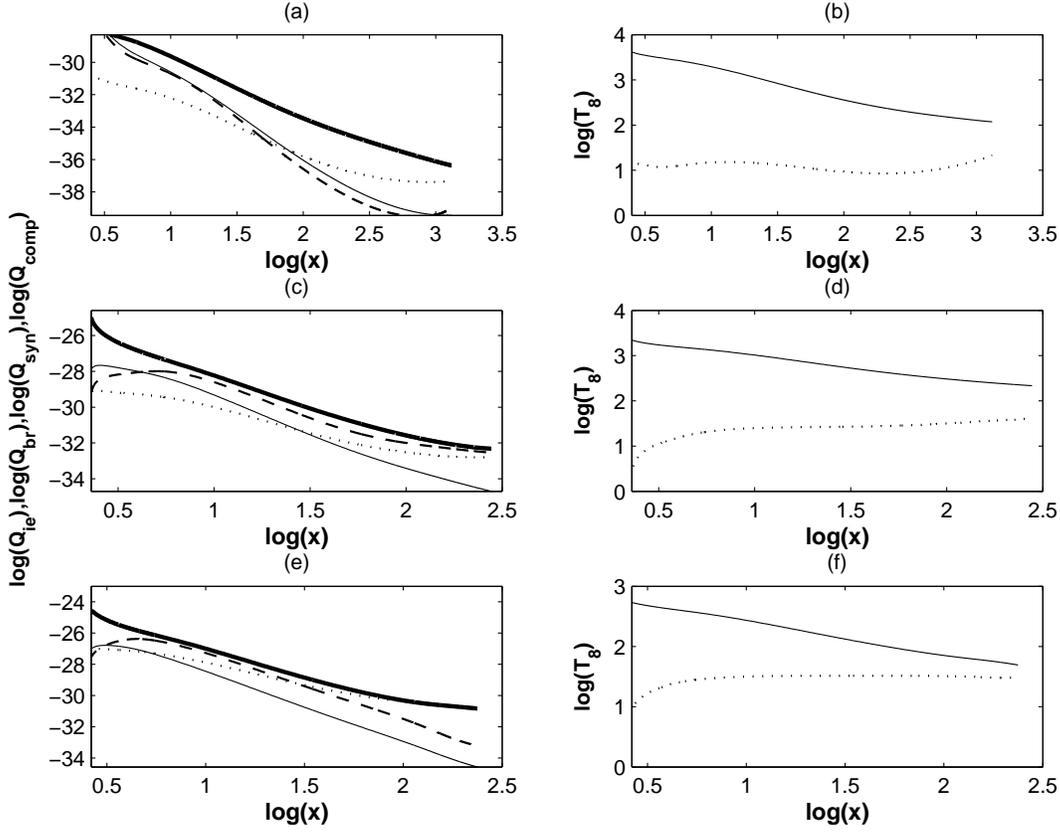}
\caption{
Variation of (a) dimensionless energy of Coulomb coupling (thicker line), 
bremsstrahlung (dotted line), synchrotron (solid), inverse Comptonization due
to synchrotron photon (dashed line) processes in logarithmic scale, 
(b) corresponding 
ion (solid) and electron (dotted) temperatures in units of $10^8$K, 
as functions of radial coordinate for $\mdot=0.01$.
(c), (e) Same as (a) except $\mdot=0.1, 1$ respectively.
(d), (f) Same as (b) except $\mdot=0.1, 1$ respectively.
Other parameters are $a=0$, $\alpha = 0.01$, $M=10$; see Table 1 for details.
}
\label{figstl0t}
\end{figure}

Figure \ref{figstl0t} describes how do the various cooling processes and corresponding 
temperature profiles vary as functions
of radial coordinate. At a low $\mdot$ ($=0.01$) the system is radiatively 
inefficient, relative to that of higher $\mdot$-s ($=0.1,1$), which brings out a hot
two temperature Keplerian-sub-Keplerian transition region. As the flow advances through the 
sub-Keplerian
part, strong two temperature nature remains intact. For a higher $\mdot$ ($=1$), 
however, the transition region is of marginally two temperatures. This is due to the
efficient bremsstrahlung radiation at high density. In the 
vicinity of black hole $T_e\sim 10^9$K in the flow with $\mdot=0.01$ when
$f\rightarrow 0$, as explained above, in the contrary to the cases with 
$\mdot=0.1,1$ when $f\rightarrow 1$ and $T_e$ sharply decreases.
Note that the accretion disc around a stellar mass black hole is
arrested by significant magnetic field. This results in dominance of the synchrotron effect
over the bremsstrahlung as the flow advances. 

\subsubsection{Kerr black holes}

\begin{figure}
\centering
\includegraphics[width=0.70\columnwidth]{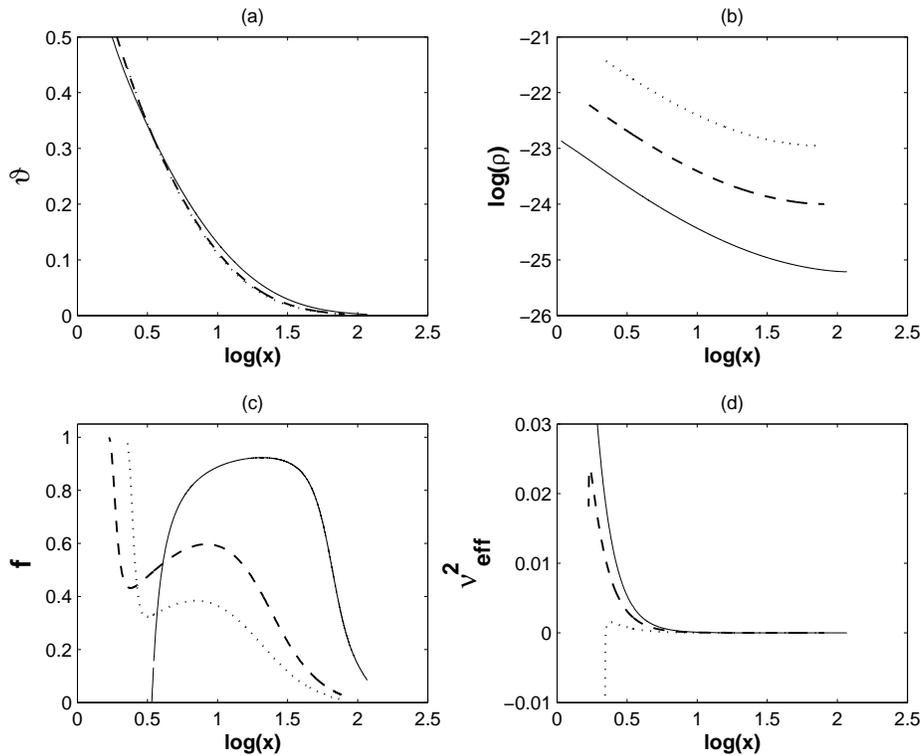}
\caption{
Same as Fig. \ref{figstl0}, except $a=0.998$.
 }
\label{figstl9} \end{figure}

We consider the rotating black holes with $a=0.998$. As discussed earlier 
(Mukhopadhyay 2003), the angular momentum of the flow should be smaller around
a rotating black hole compared to that around a static black hole. This reassembles
advancing the Keplerian component which decreases size of the sub-Keplerian
part. As the disc remains Keplerian (which is radiatively efficient) upto, 
e.g., $x\sim 100$ (see the outer radius 
in Fig. \ref{figstl9}), the flow cools down significantly
before deviating to the sub-Keplerian regime. Figure \ref{figstl9}a shows
that the velocity profiles for all $\mdot$-s are similar to each other,
in absence of a strong centrifugal barrier. However, $f$, while very small
at $x\sim 100$, increases as the sub-Keplerian flow advances. This is 
because the residence time of the flow decreases in an element of the sub-Keplerian disc
hindering cooling processes to complete. Figure \ref{figstl9}c 
shows that even $f\rightarrow 0.9$ 
at $x\sim 30$ for $\mdot=0.01$, when the density is lowest (see Fig. \ref{figstl9}b).

\begin{figure}
\centering
\includegraphics[width=0.80\columnwidth]{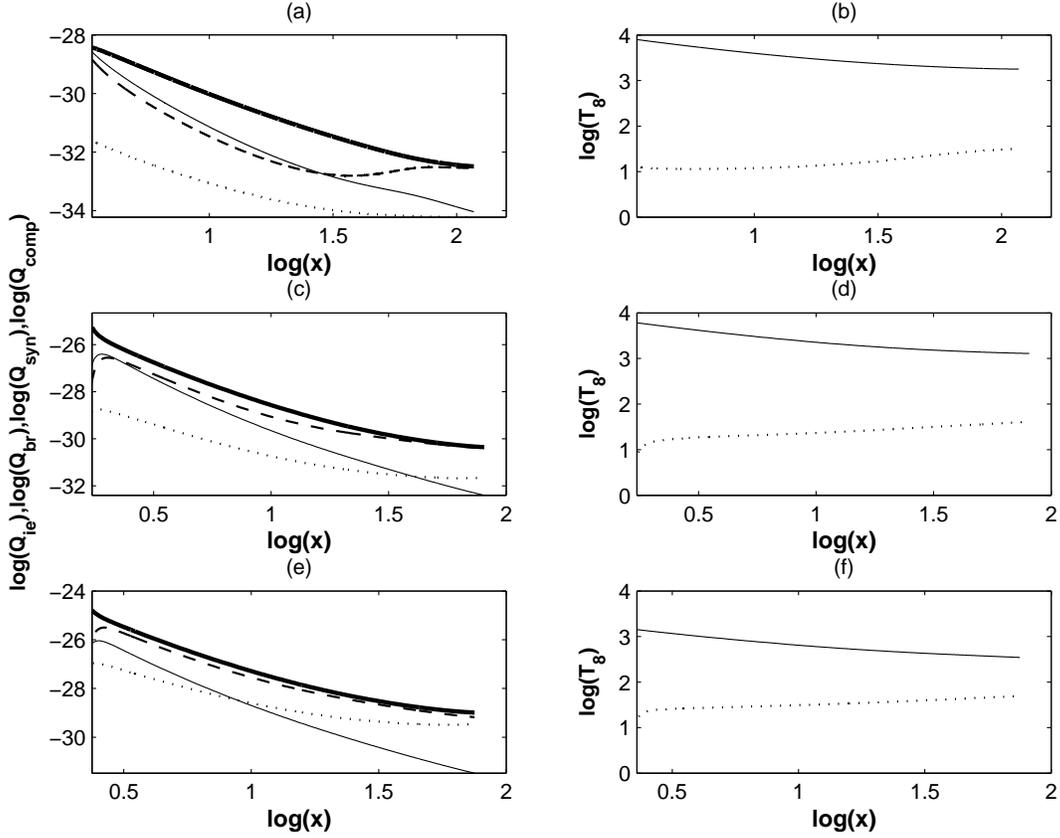}
\caption{
Same as Fig. \ref{figstl0t}, except $a=0.998$.
}
\label{figstl9t}
\end{figure}

However, as the flow approaches to the black hole the synchrotron emission increases, 
and hence the system acquires enough soft photons 
which help in occurring the inverse Compton process. As a result
the flow cools down further. When $\mdot=0.01$ the cooling process at the very inner edge 
of the accretion disc is dominant due to relatively high residence time of 
the flow, compared to that of a higher $\mdot$, rendering high $T_e$ and then $f\rightarrow 0$
very close to the black hole. Figure \ref{figstl9}d proves that the
flow is convectionally stable all the way upto the black hole event horizon. 
For $\mdot\gsim 0.1$, on the other hand, 
the flow is strongly advective and then unable to cool
down before plunging into the black hole.
Figure \ref{figstl9t} shows the profiles of cooling processes and ion-electron 
temperatures. Basic nature of the profiles is pretty similar to that of Schwarzschild
cases, except all of them advance in. 

\subsection{Super-Eddington accretors}

The ``radiation trapped'' accretion disc can be attributed to 
the radiatively driven outflow or jet. 
This is likely to occur when the accretion rate is super-Eddington
(Lovelace et al. 1994, Begelman et al. 2006, Fabbiano 2004, Ghosh \& 
Mukhopadhyay 2009),
as seen in the ultra luminous X-ray (ULX) sources such as SS433 (with luminosity 
$\sim 10^{40}$ erg/s or so; Fabrika 2004). In order to describe such sources, the
models described below are the meaningful candidates. We consider
$\mdot=10,100$.

\subsubsection{Schwarzschild black holes}

\begin{figure}
\centering
\includegraphics[width=0.70\columnwidth]{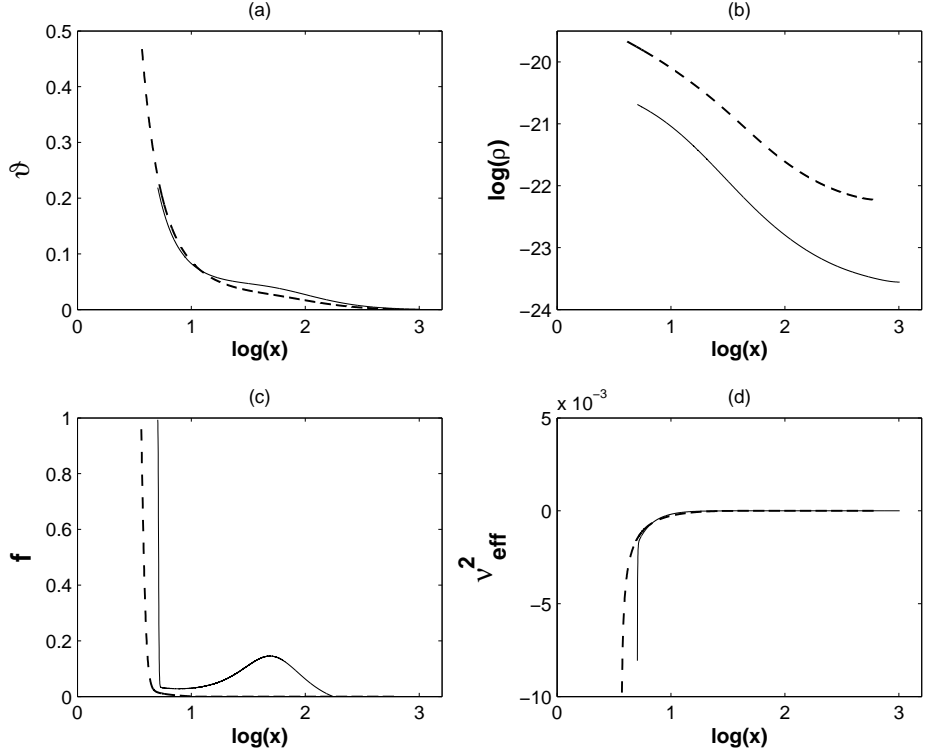}
\caption{
Variation of dimensionless (a) radial velocity, (b) density,
(c) cooling factor,
(d) square of convective frequency, as functions of radial 
coordinate for super-Eddington accretion flows.
Solid, dashed curves are for $\mdot=10,100$ respectively.
Other parameters are $a=0$, $\alpha = 0.01$, $M=10$; see Table 1 for details.
 }
\label{figsth0} 
\end{figure}

\begin{figure}
\centering
\includegraphics[width=0.70\columnwidth]{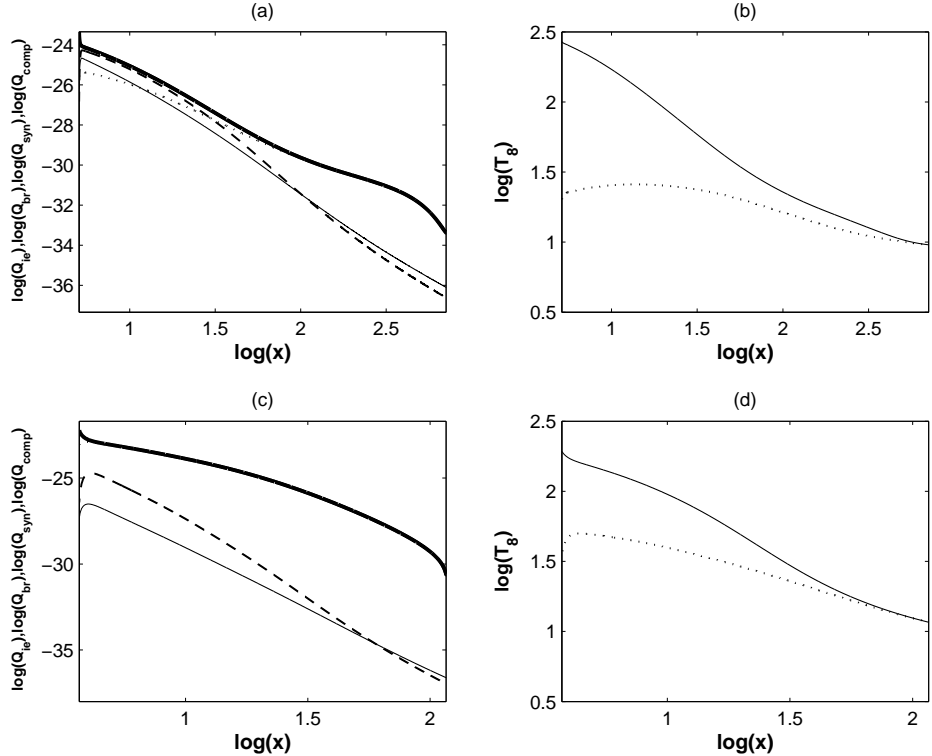}
\caption{
Variation of (a) dimensionless energy of Coulomb coupling (thicker line), 
bremsstrahlung (dotted line), synchrotron (solid), inverse Comptonization due
to synchrotron photon (dashed line) processes in logarithmic scale, 
(b) corresponding 
ion (solid) and electron (dotted) temperatures in units of $10^8$K, 
as functions of radial coordinate for $\mdot=10$.
(c) Same as (a) except $\mdot=100$.
(d) Same as (b) except $\mdot=100$.
Other parameters are $a=0$, $\alpha = 0.01$, $M=10$; see Table 1 for details.
}
\label{figsth0t}
\end{figure}

A high mass accretion rate significantly enhances density, upto two orders
of magnitude compared to that of a low $\mdot$, which severely affects $f$
and finally temperature profiles. The profiles of velocity shown in
Fig. \ref{figsth0} are quite similar to that of sub-Eddington and
Eddington cases. Because
of similar reasons explained in \S3.1.1 the profile exhibits a stronger centrifugal
barrier for a lower $\mdot$ ($=10$). A lower $\mdot$ flow will have relatively more
gas and then quasi-spherical structure compared to that of a higher $\mdot$ ($=100$), 
which results in a lower velocity in the latter case. On the other hand,
a lower velocity corresponds to a higher density which results in the strong
bremsstrahlung radiation rendering $f\rightarrow 0$. For a lower
$\mdot$, at $x\sim 50$, the energy radiated due to bremsstrahlung process
becomes weaker than the energy transferred from protons to electrons
through the Coulomb coupling (see Fig. \ref{figsth0t}), which increases $f$
(see Fig. \ref{figsth0}c). Subsequently, the synchrotron
process becomes dominant (see Fig. \ref{figsth0t}), reassembling 
$f\rightarrow 0$. However, very
close to the black hole a strong advection does not allow the flow, independent
of $\mdot$, to radiate efficiently rendering $f\rightarrow 1$ again.
This also results in marginal convective instability at $x<10$, as shown
in Fig. \ref{figsth0}d.

Figure \ref{figsth0t} shows that discs remain of one temperature at 
the transition radius. As the flows advance with a sub-Keplerian angular
momentum, $T_p$ profile deviates from that of $T_e$. For $\mdot=100$,
the high density flow is dominated by very efficient bremsstrahlung 
radiation all the way. In the vicinity of the black hole, an efficient
cooling reassembles a sharp downfall of $T_e$.

\subsubsection{Kerr black holes}

\begin{figure}
\centering
\includegraphics[width=0.70\columnwidth]{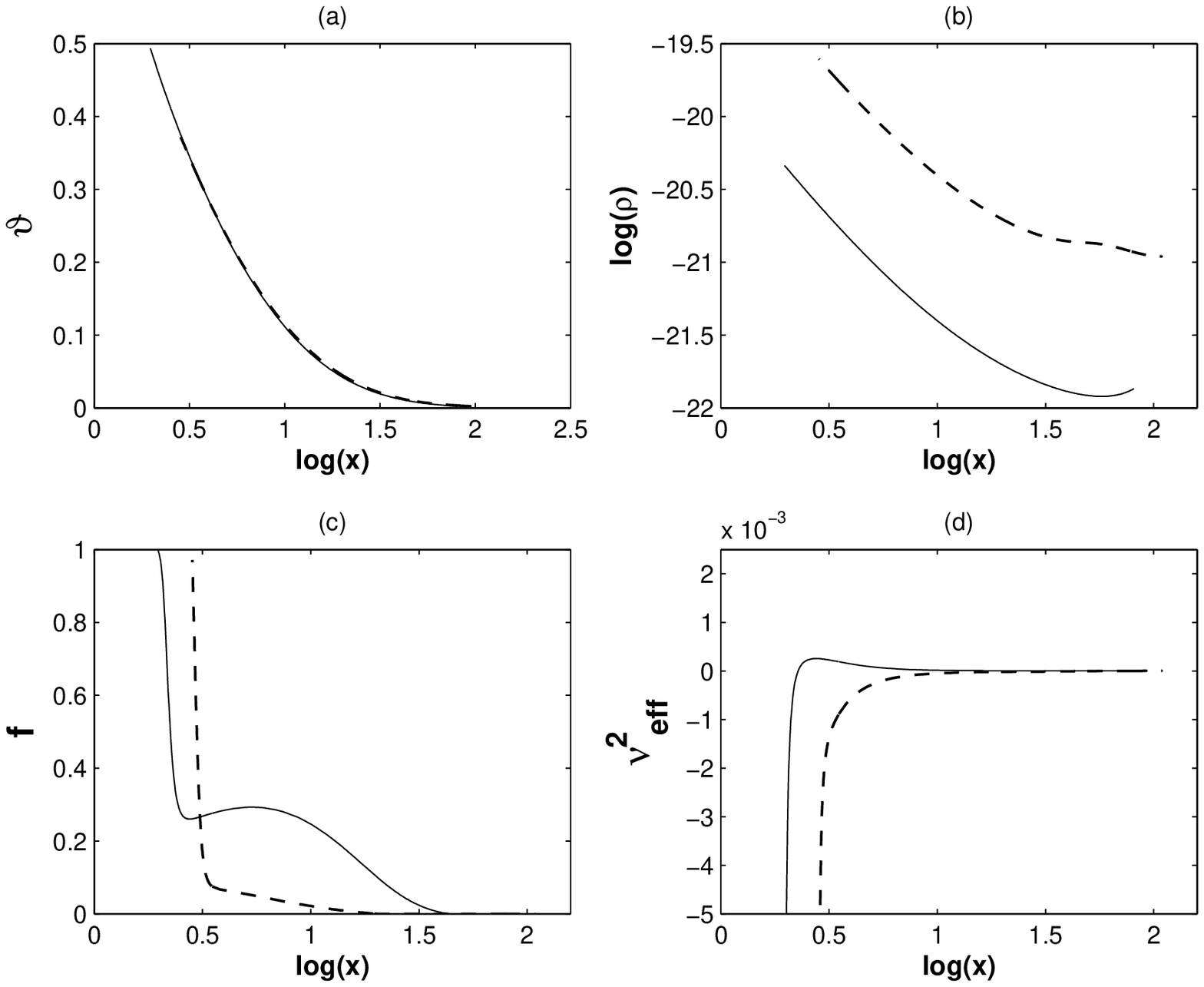}
\caption{
Same as Fig. \ref{figsth0}, except $a=0.998$.
 }
\label{figsth9} 
\end{figure}

As $\lambda$ decreases in the case of a higher $a$, 
like low $\mdot$ cases (see Mukhopadhyay 2003), the
transition region advances an order of magnitude
compared to that of Schwarzschild black holes. Similar to the 
cases of low $\mdot$, as shown in Fig. \ref{figsth9}a, 
any centrifugal barrier smears out. However, unlike the flow 
around a static black hole, here the disc with $\mdot=10$ remains
stable upto very close to the black hole. The reason is that
a high $a$ corresponds to a larger inner edge of the disc and thus
the residence time of matter in the disc is higher. 
As a result the radiative processes keep cooling 
and then stabilizing the flow upto very inner edge.

\begin{figure}
\centering
\includegraphics[width=0.70\columnwidth]{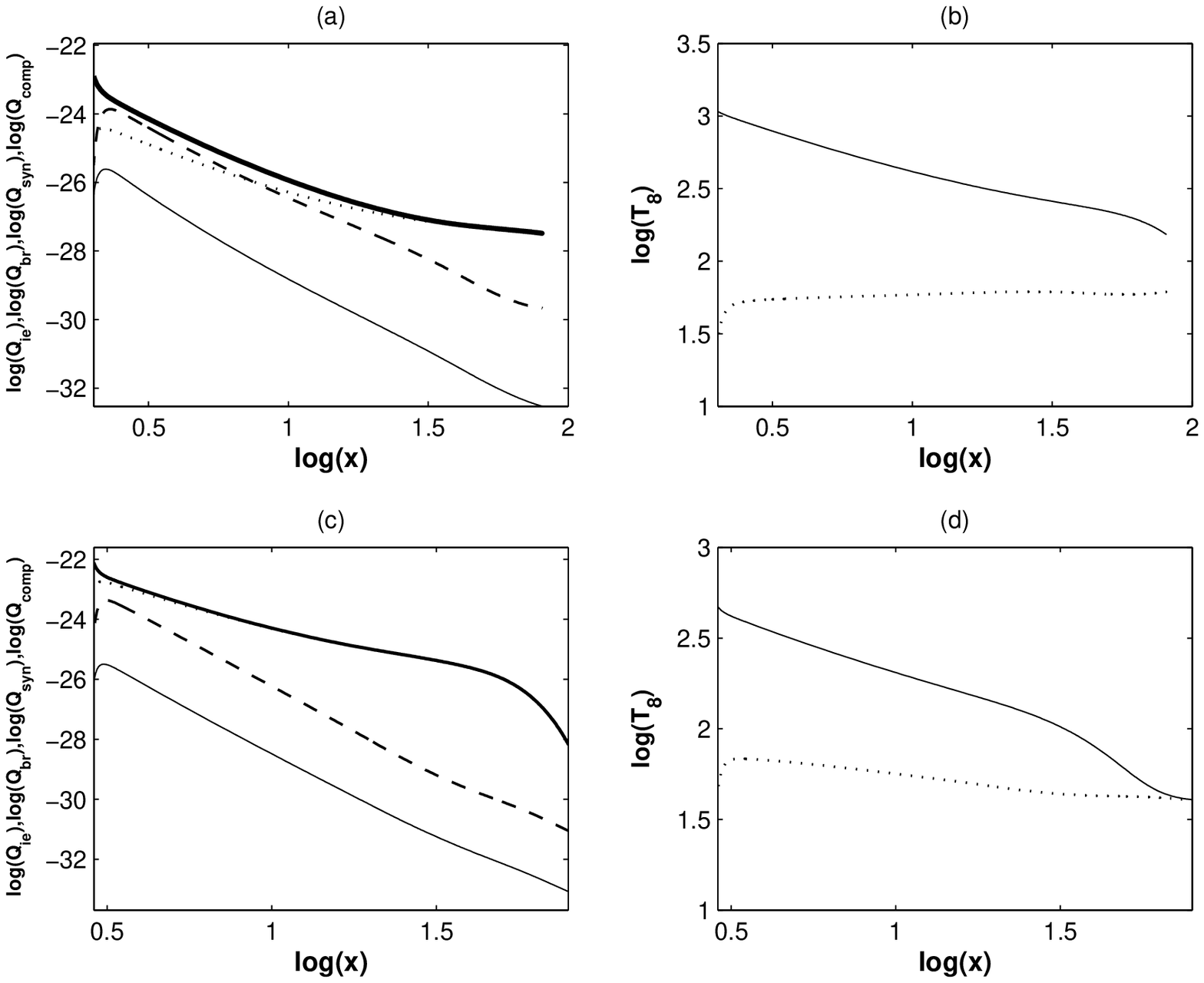}
\caption{
Same as Fig. \ref{figsth0t}, except $a=0.998$.
}
\label{figsth9t}
\end{figure}

Figure \ref{figsth9t} shows that although a high $\mdot$ exhibits
a one temperature transition zone due to extremely efficient cooling processes,
particularly due to bremsstrahlung radiation, as $\mdot$ decreases
the Keplerian disc itself becomes of two temperatures 
before deviating to the sub-Keplerian zone, unlike that of the Schwarzschild case. 
This is mainly because
a flow with a high $a$ brings the Keplerian disc further in where the
transport of angular momentum increases leading to the decrease of
the residence time of matter which does not allow an efficient cooling.
However, the basic behaviours of various cooling processes is
pretty similar to that around a static black hole.

\section{Two temperature accretion disc around supermassive black holes}

As of stellar mass black holes, here also we concentrate on two extreme regimes: 
(1) sub-Eddington and 
Eddington limits of accretion, (2) super-Eddington accretion; focusing on both
nonrotating (Schwarzschild) and rotating 
(Kerr with $a=0.998$) black holes. 

\subsection{Sub-Eddington and Eddington accretors}

The under-luminous AGNs and quasars (e.g. Sgr~$A^{*}$)
had been already described by advection dominated model,
where the flow is expected to be substantially sub-critical/sub-Eddington
with a very low luminosity ($\lsim 10^{35}$ erg/s).
Therefore the present cases, particularly of $\mdot\lsim 0.01$,
could be potential models in order to describe under-luminous
sources.

\subsubsection{Schwarzschild black holes}

\begin{figure}
\centering
\includegraphics[width=0.70\columnwidth]{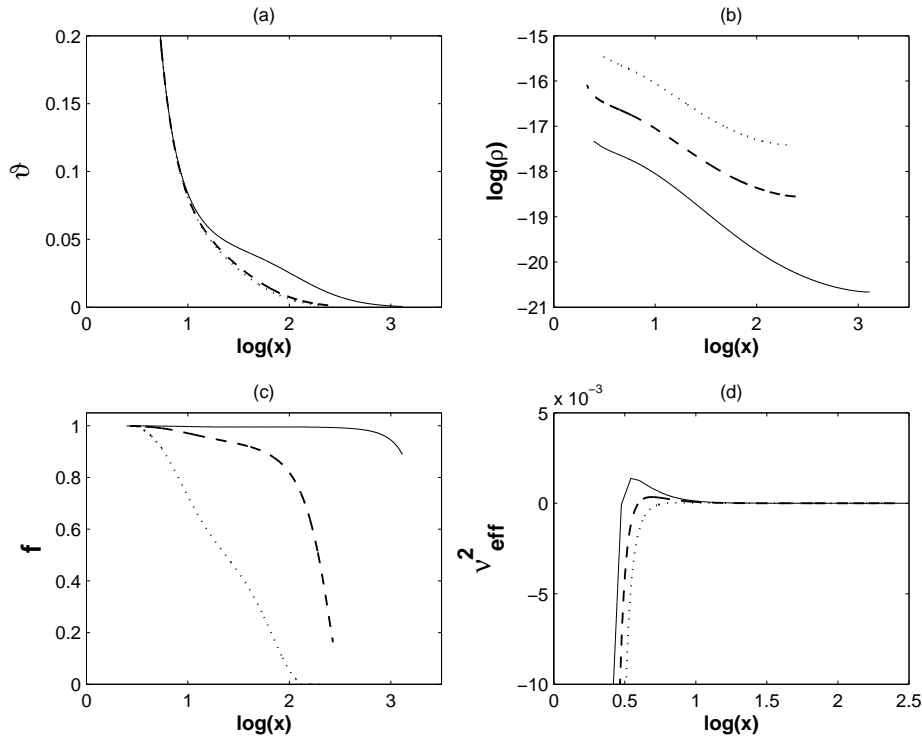}
\caption{
Variation of dimensionless (a) radial velocity, (b) density,
(c) cooling factor,
(d) square of convective frequency, as functions of radial 
coordinate for sub-Eddington and Eddington accretion flows.
Solid, dashed, dotted curves are for $\mdot=0.01,0.1,1$ respectively.
Other parameters are $a=0$, $\alpha = 0.01$, $M=10^7$; see Table 2 for details.
 }
\label{figsul0} 
\end{figure}

\begin{figure}
\centering
\includegraphics[width=0.80\columnwidth]{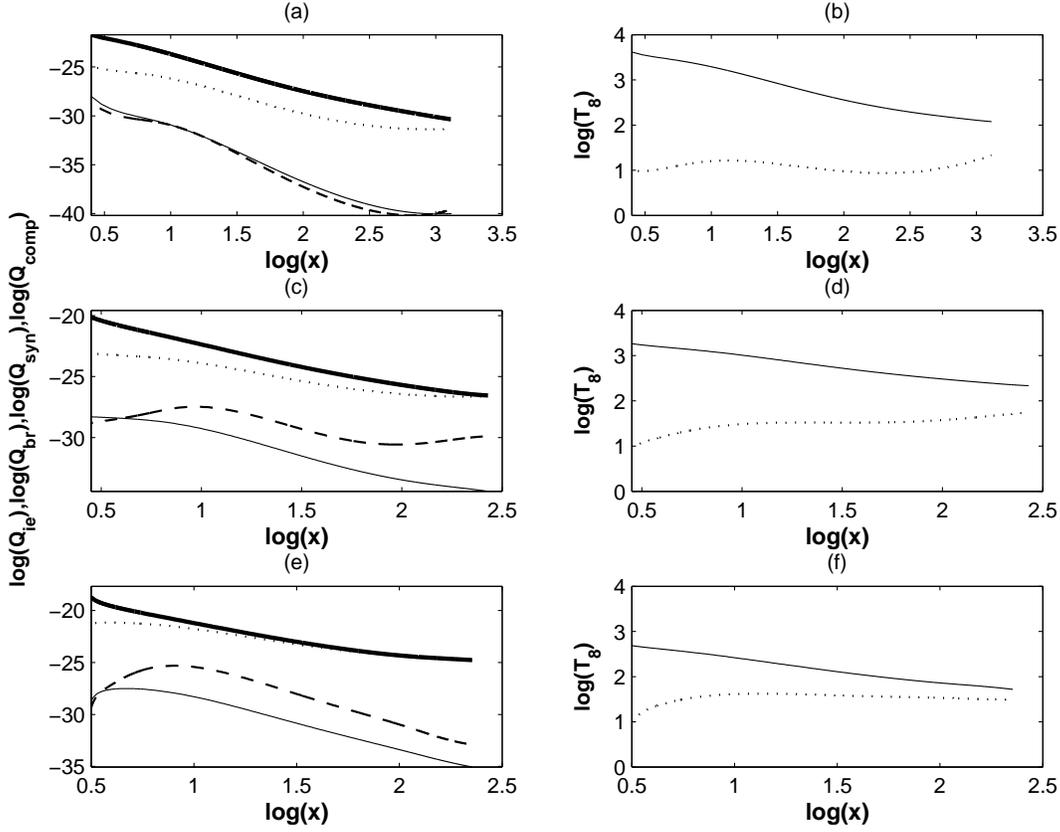}
\caption{
Variation of (a) dimensionless energy of Coulomb coupling (thicker line), 
bremsstrahlung (dotted line), synchrotron (solid), inverse Comptonization due
to synchrotron photon (dashed line) processes in logarithmic scale, 
(b) corresponding 
ion (solid) and electron (dotted) temperatures in units of $10^8$K, 
as functions of radial coordinate for $\mdot=0.01$.
(c), (e) Same as (a) except $\mdot=0.1, 1$ respectively.
(d), (f) Same as (b) except $\mdot=0.1, 1$ respectively.
Other parameters are $a=0$, $\alpha = 0.01$, $M=10^7$; see Table 2 for details.
}
\label{figsul0t}
\end{figure}

Table 2 lists the sets of input parameters for the model cases described here.
Naturally a disc around a supermassive black hole will have much lower
density compared to that around a stellar mass black hole. Therefore, the cooling
processes, particularly the bremsstrahlung radiation which is density 
dependent, are expected to be inefficient leading to a high $f$.
However, the velocity profiles shown in Fig. \ref{figsul0}a are very
similar/same to that around a stellar mass black hole.
Figure \ref{figsul0}c shows that $f\rightarrow 1$ in most of
the sub-Keplerian regime for $\mdot=0.01$. As $\mdot$ increases, the density
increases and thus the bremsstrahlung radiation increases, as shown in 
Fig. \ref{figsul0t}, which leads to the transition of radiatively inefficient flow
to GAAF. When $\mdot=1$ the bremsstrahlung effect is very high resulting an 
GAAF with $f$ much smaller than unity upto very close to the black hole.
Figure \ref{figsul0}d shows that close to the black hole 
there is a possible convective instability for all $\mdot$-s.
This is because a strong advection of matter close to the 
black hole hindering cooling processes which results in $f\rightarrow 1$.
This reassembles a possible convective instability at the inner edge. For a higher $\mdot$,
the density is high which favours convection and thus brings the
convective instability earlier in, at a relatively outer radius.

\newpage
\noindent{ Table 2: Parameters for accretion with $\alpha=0.01$ 
around black holes of $M=10^7$, when the subscript `c' indicates the quantity at the
critical radius and $T_{ec}$ is expressed in units of $m_i c^2/k$ }

\begin{center}
\begin{tabular}{lllllllllllll}
\hline
\hline
$\mdot$ & $a$ & $\gamma$ & $x_{c}$ & $\lambda _{c}$ & $T_{ec}$  \\
\hline
\hline
& & Sub-Eddington, & Eddington & accretors & \\
\hline
\hline
0.01 & 0     & 1.5  & 5.5 & 3.2 & 0.0001 \\
0.01 & 0.998 & 1.5  & 3.5 & 1.7 & 0.0001 \\
\hline
0.1  & 0     & 1.4  & 5.5 & 3.2 & 0.000178\\
0.1  & 0.998 & 1.4  & 3.5 & 1.7 & 0.00023\\
\hline
1    & 0     & 1.35 & 5.5 & 3.2 & 0.0002493\\
1    & 0.998 & 1.35 & 3.5 & 1.7 & 0.000295\\
\hline
\hline
& & Super-Eddington & accretors & & \\
\hline
\hline
10   & 0     & 1.345& 5.5 & 3.2 & 0.000427\\
10   & 0.998 & 1.345& 3.5 & 1.7 & 0.0006\\
\hline
100  & 0     & 1.34 & 5.5 & 3.2 & 0.0003874\\
100  & 0.998 & 1.34 & 3.5 & 1.7 & 0.00059\\
\hline
\hline
\end{tabular}
\end{center}

The temperature profiles shown in Fig. \ref{figsul0t} are pretty similar
to what we obtain in stellar mass black holes what we do not explain here
again. However, note that unlike stellar mass black holes, only the
bremsstrahlung radiation is effective in cooling the flow around a
supermassive black hole, particularly for $\mdot=1$. 

\subsubsection{Kerr black holes}

\begin{figure}
\centering
\includegraphics[width=0.70\columnwidth]{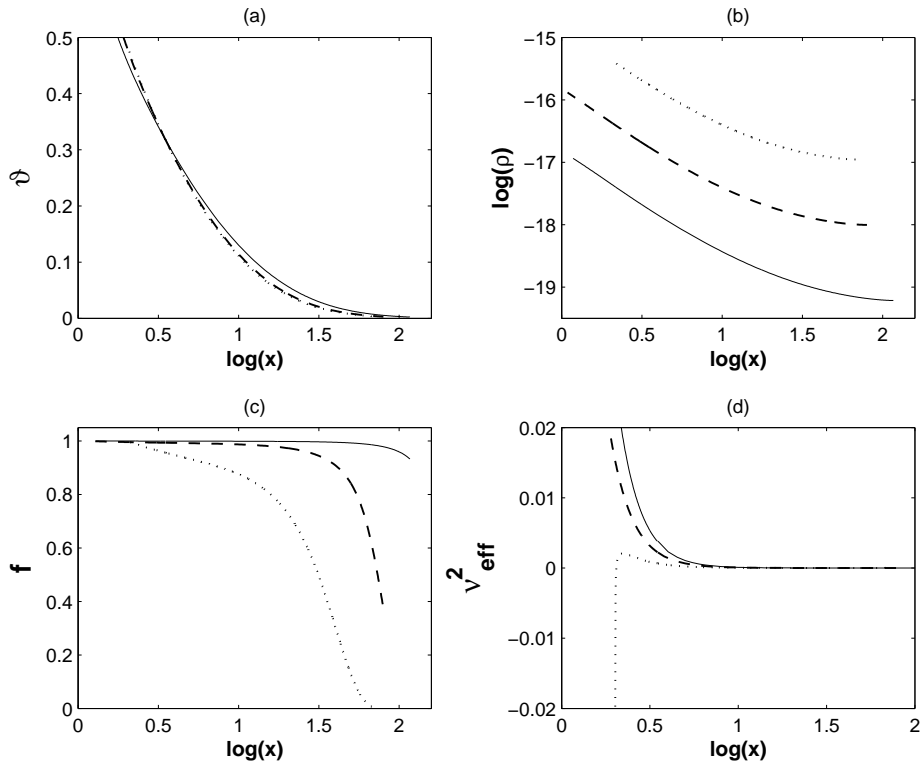}
\caption{
Same as Fig. \ref{figsul0}, except $a=0.998$.
 }
\label{figsul9} \end{figure}

The specific angular momentum of the black hole is chosen to be $a=0.998$.
The basic hydrodynamical properties are similar to that in Schwarzschild cases,
except, like the flows around stellar mass black holes, the transition region
advances due to a smaller angular momentum of the flow, shown in Fig. \ref{figsul9}. 
As discussed in \S3.2.2, a higher $a$ corresponds to a smaller
$\lambda$ which in turn decreases $\vartheta$ at a particular radius of
the inner edge of the disc, when the inner edge is stretched in, compared
to that around a Schwarzschild black hole. This results in the increase of residence 
time of the flow in the sub-Keplerian disc before plunging into the
black hole. Therefore, the bremsstrahlung
process keeps cooling and then stabilizing the flow upto very inner edge. 
Important point to note, as a consequence, is that the disc flow around a rotating black hole
is convectively more stable compared to that around a nonrotating black hole.
This is particularly because the density gradient of the inner 
(e.g. the vicinity of $x=2$, which is the event horizon for a nonrotating black hole) 
flow around a rotating 
black hole is less stepper compared to that around
a nonrotating black hole, and hence the flow is convectively more stable in
the former case.

\begin{figure}
\centering
\includegraphics[width=0.80\columnwidth]{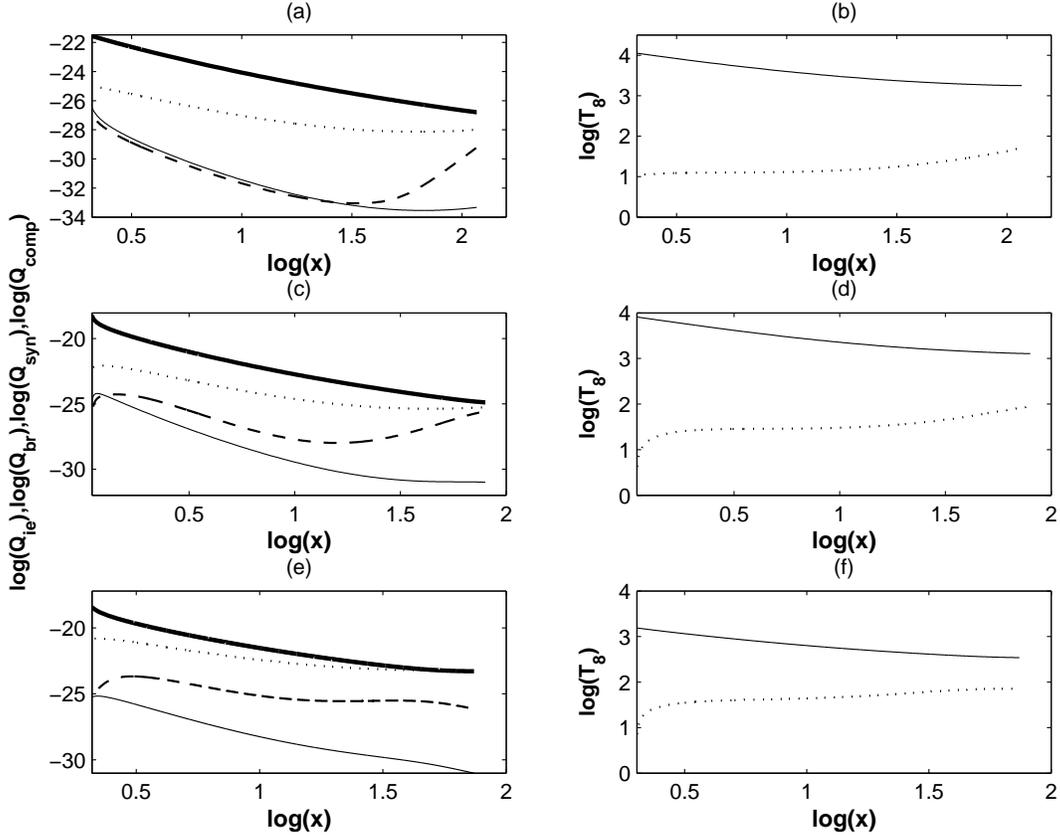}
\caption{
Same as Fig. \ref{figsul0t}, except $a=0.998$.
}
\label{figsul9t}
\end{figure}

Figure \ref{figsul9t} shows that basic features of the temperature profiles
are similar to the cases of static black holes. However, the transition region
reveals that for a lower $\mdot$, the Keplerian flow exhibits the
inverse Comptonization via synchrotron photons.
As the flow advances
with a sub-Keplerian angular momentum, the residence time decreases and thus
inverse Comptonization decreases, resulting a hotter flow.

\subsection{Super-Eddington accretors}

Ultra-luminous accretors with a high kinetic luminosity
($\sim 10^{46} - 10^{49}$ erg/s)
radio jet have been observed in the highly luminous AGNs and ultra-luminous quasars 
(e.g. PKS~0743-67; Punsly \& Tingay 2005), possibly in ULIRs 
(Genzel et al. 1998) and narrow-line Seyfert 1 galaxies (e.g. Mineshige et al. 2000).
Therefore, the following cases could be potential models to 
explain such sources.

\subsubsection{Schwarzschild black holes}

\begin{figure}
\centering
\includegraphics[width=0.70\columnwidth]{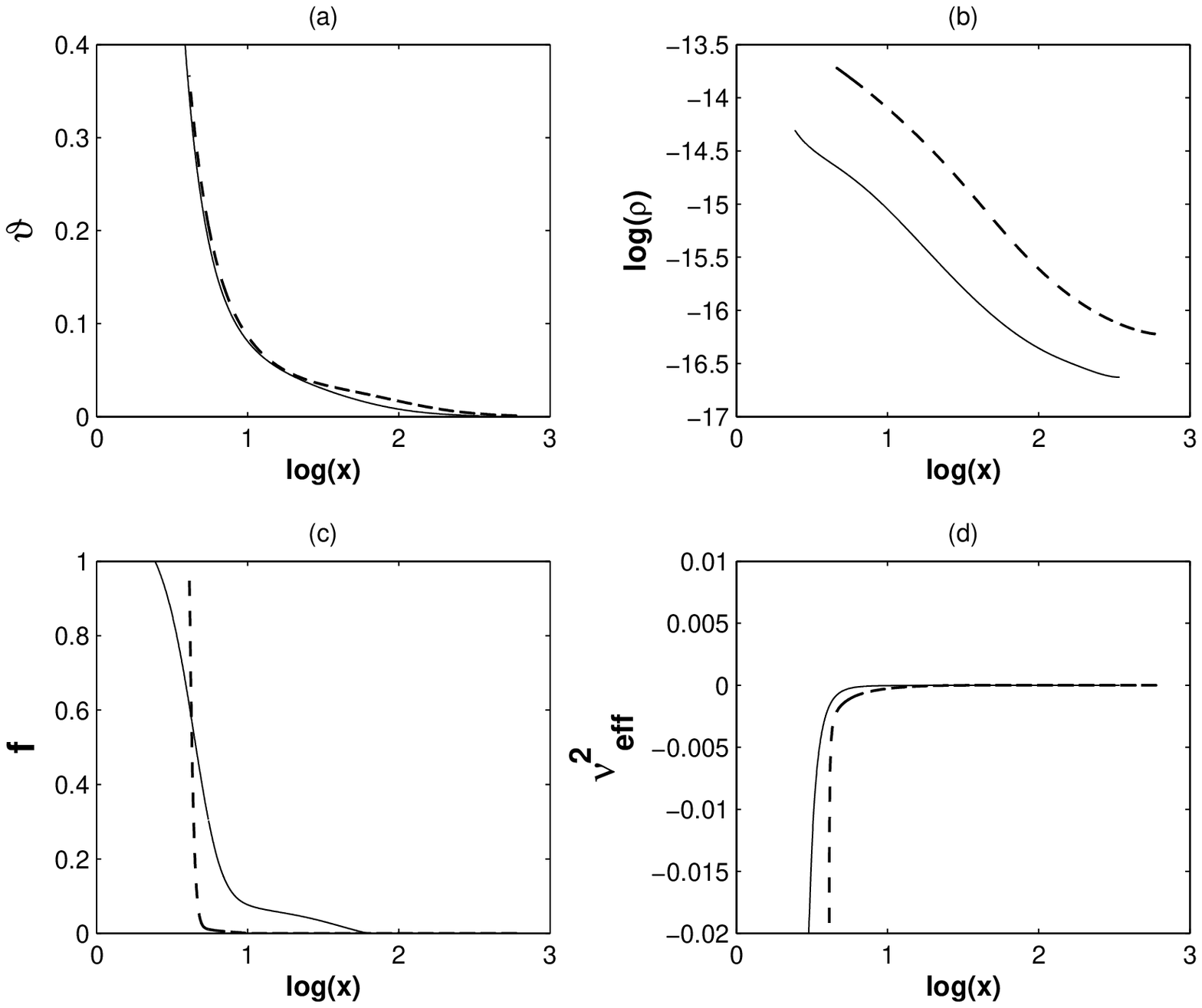}
\caption{
Variation of dimensionless (a) radial velocity, (b) density,
(c) cooling factor,
(d) square of convective frequency, as functions of radial 
coordinate for super-Eddington accretion flows.
Solid, dashed curves are for $\mdot=10,100$ respectively.
Other parameters are $a=0$, $\alpha = 0.01$, $M=10^7$; see Table 2 for details.
 }
\label{figsuh0} 
\end{figure}

\begin{figure}
\centering
\includegraphics[width=0.70\columnwidth]{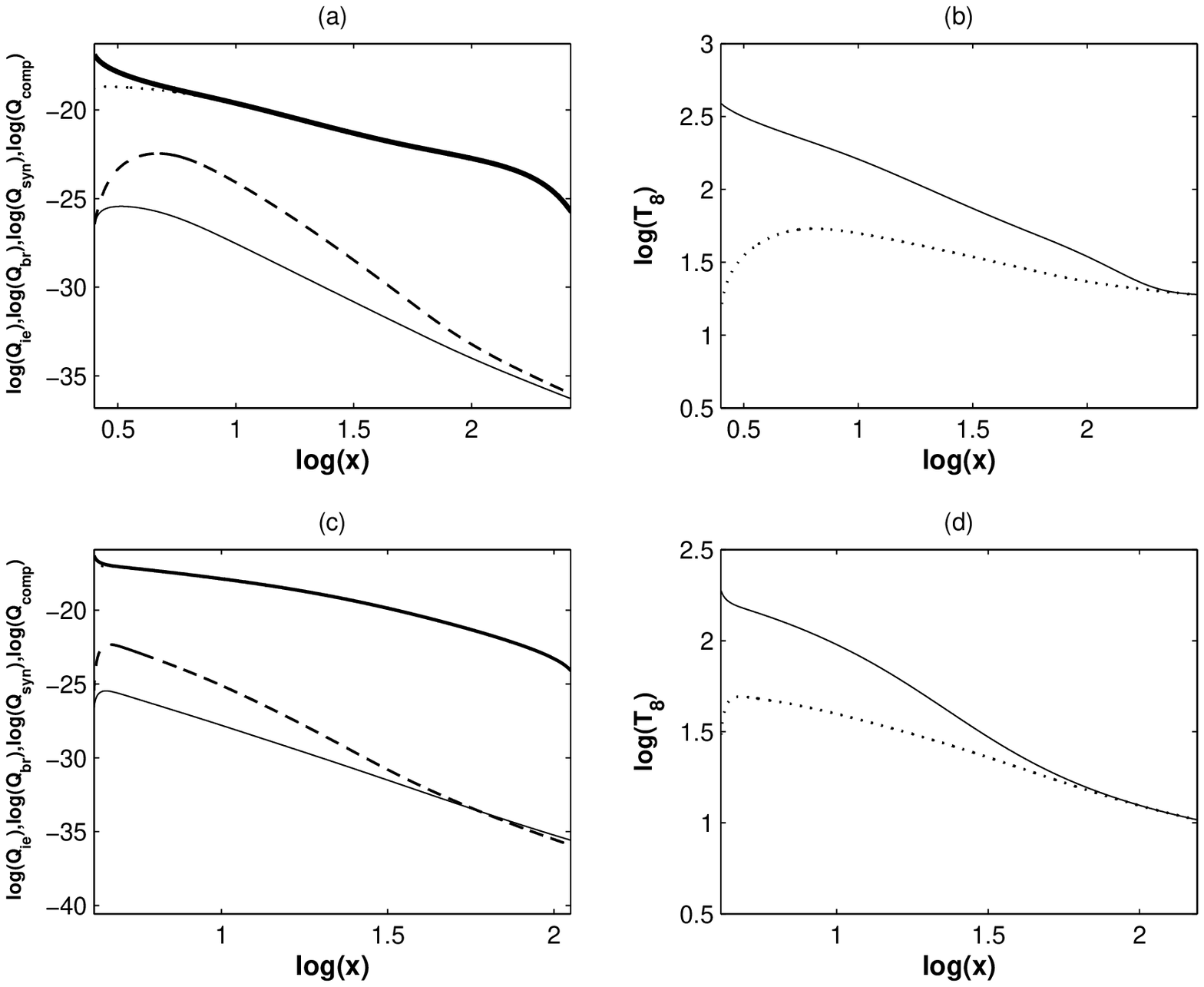}
\caption{
Variation of (a) dimensionless energy of Coulomb coupling (thicker line), 
bremsstrahlung (dotted line), synchrotron (solid), inverse Comptonization due
to synchrotron photon (dashed line) processes in logarithmic scale, 
(b) corresponding 
ion (solid) and electron (dotted) temperatures in units of $10^8$K, 
as functions of radial coordinate for $\mdot=10$.
(c) Same as (a) except $\mdot=100$.
(d) Same as (b) except $\mdot=100$.
Other parameters are $a=0$, $\alpha = 0.01$, $M=10^7$; see Table 2 for details.
}
\label{figsuh0t}
\end{figure}

Figures \ref{figsuh0} and \ref{figsuh0t} show that the basic 
flow properties are pretty similar to that around stellar
mass black holes, except in the present cases the centrifugal
barrier smears out. This is because a high black hole mass
corresponds to a low density of the flow and thus a fast infall.
This also results, unlike stellar mass black holes, in an inefficient
synchrotron radiation even at the inner edge of the disc. 

\subsubsection{Kerr black holes}

\begin{figure}
\centering
\includegraphics[width=0.70\columnwidth]{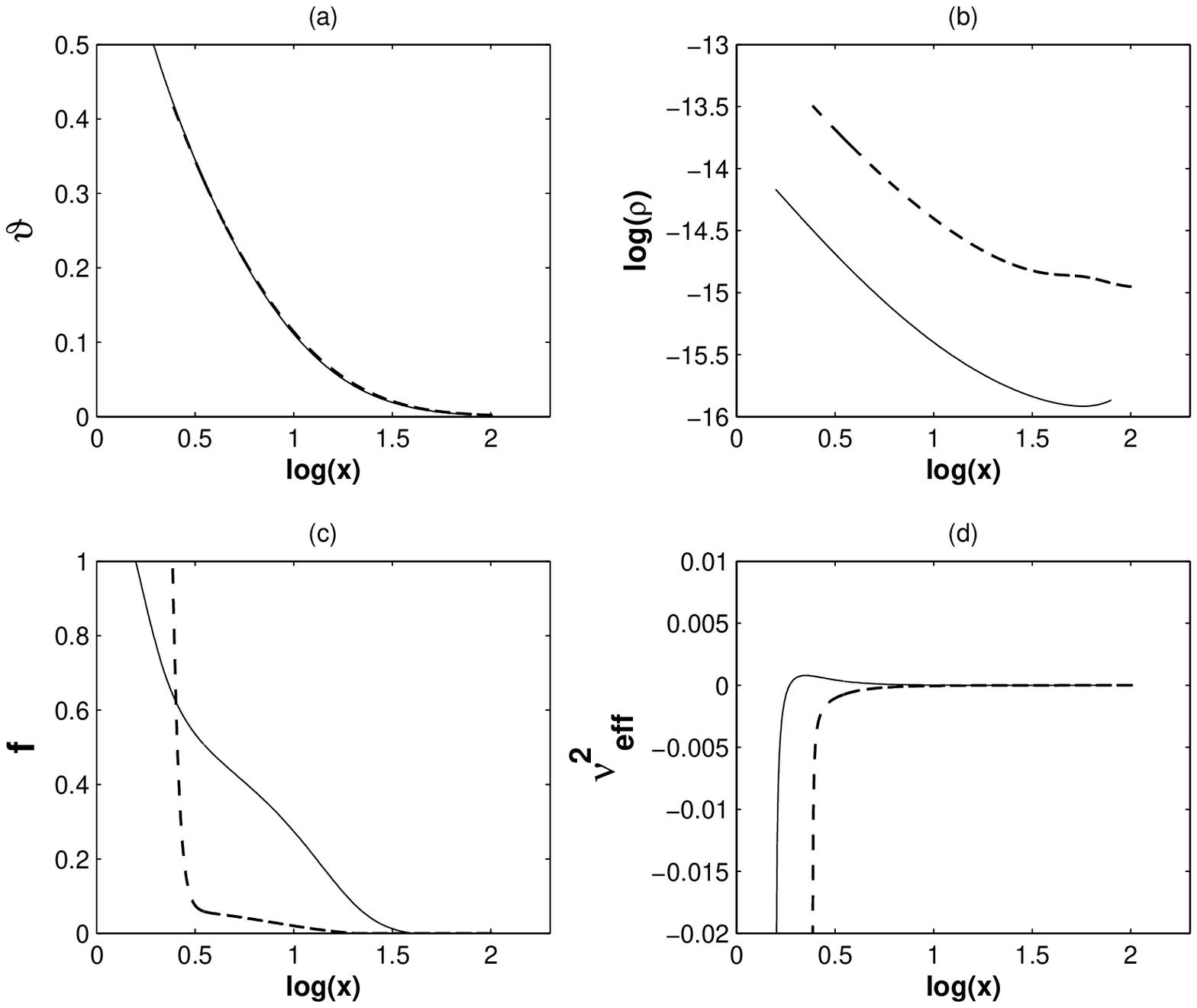}
\caption{
Same as Fig. \ref{figsuh0}, except $a=0.998$.
 }
\label{figsuh9} 
\end{figure}

\begin{figure}
\centering
\includegraphics[width=0.70\columnwidth]{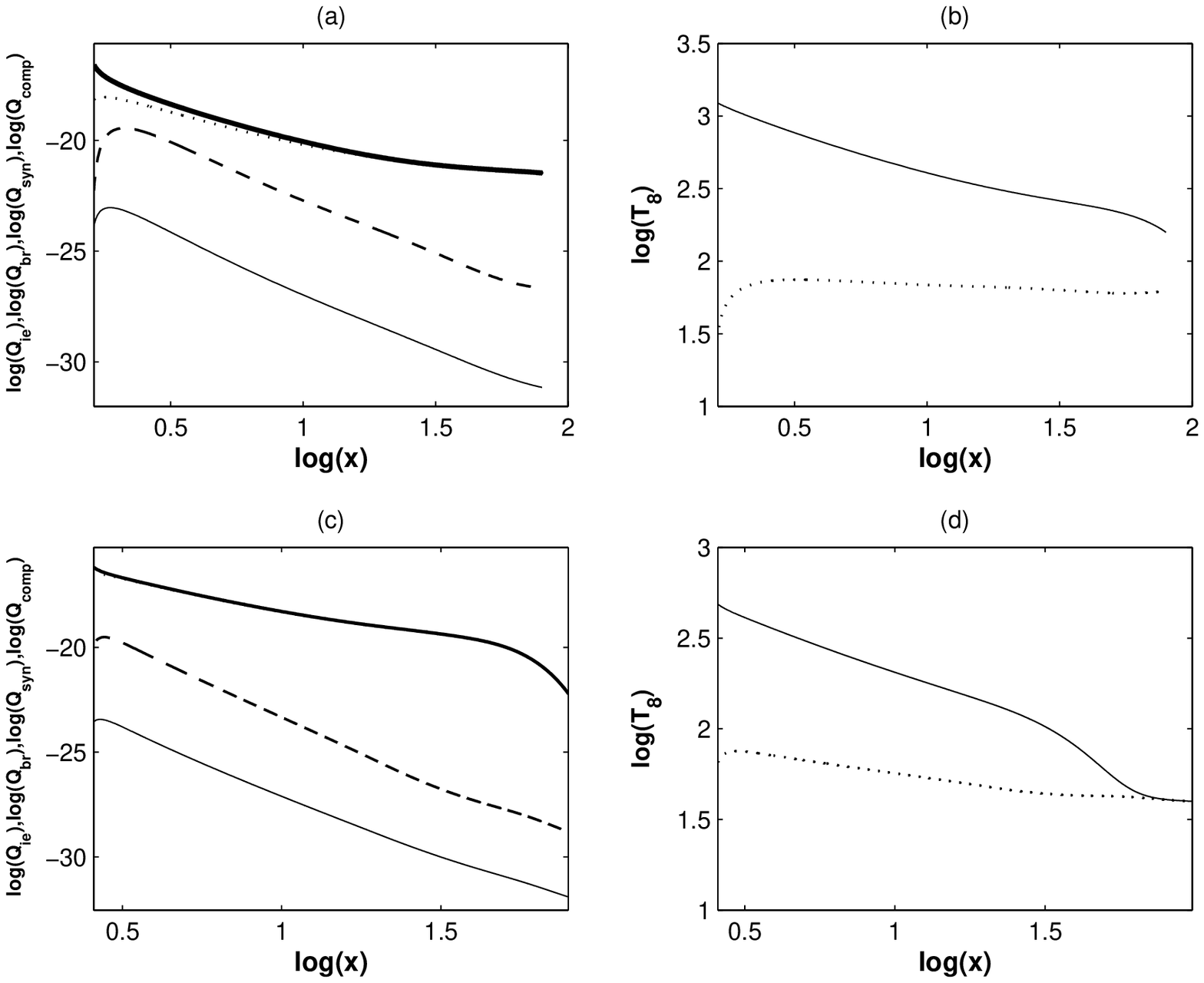}
\caption{
Same as Fig. \ref{figsuh0t}, except $a=0.998$.
}
\label{figsuh9t}
\end{figure}

Figures \ref{figsuh9} and \ref{figsuh9t} repeat the same story of
that around stellar mass black holes, but with the smeared  
centrifugal barrier, as described above for static black holes.  
However, due to decreasing density, the overall cooling effects decrease
keeping the disc hotter, particularly for $\mdot=10$.

\section{Comparison between flows with different {\huge $\alpha$} and around corotating
and counter rotating black holes}

So far we have restricted to a typical Shakura-Sunyaev viscosity parameter $\alpha=0.01$,
for corotating black holes. Now we plan to explore a lower $\alpha$, as well
as a counter rotating black hole to understand any significant change in the flow behaviour.

\subsection{Comparison between flows with $\alpha=0.01$ and $\alpha=0.0001$}

Decreasing $\alpha$ naturally decreases the rate of energy-momentum transfer 
between any two successive layers of the fluid element and increasing the residence
time of the flow in the sub-Keplerian disc. This also recedes 
the Keplerian-sub-Keplerian transition region further out. This is mainly
because a low value of $\alpha$ can not keep continuing the outward
angular momentum transport efficiently in the Keplerian flow 
below a certain radial coordinate. Therefore, the disc flow can not
remain Keplerian and becomes sub-Keplerian at a larger radius, compared 
to a flow of high $\alpha$.

We know, on the other hand, that increasing residence time 
increases the possibility of completing various radiative processes
in the disc flow, before the infalling matter plunges into the black hole.
Therefore, the flow is expected to appear cooler with smaller $f$.
Hence, for the purpose of comparison, 
we consider a flow with $\mdot=0.01$
around a supermassive black hole of $M=10^7$, e.g. Sgr~$A^*$,
which is radiatively inefficient and hot for $\alpha=0.01$.

\begin{figure}
\centering
\includegraphics[width=0.70\columnwidth]{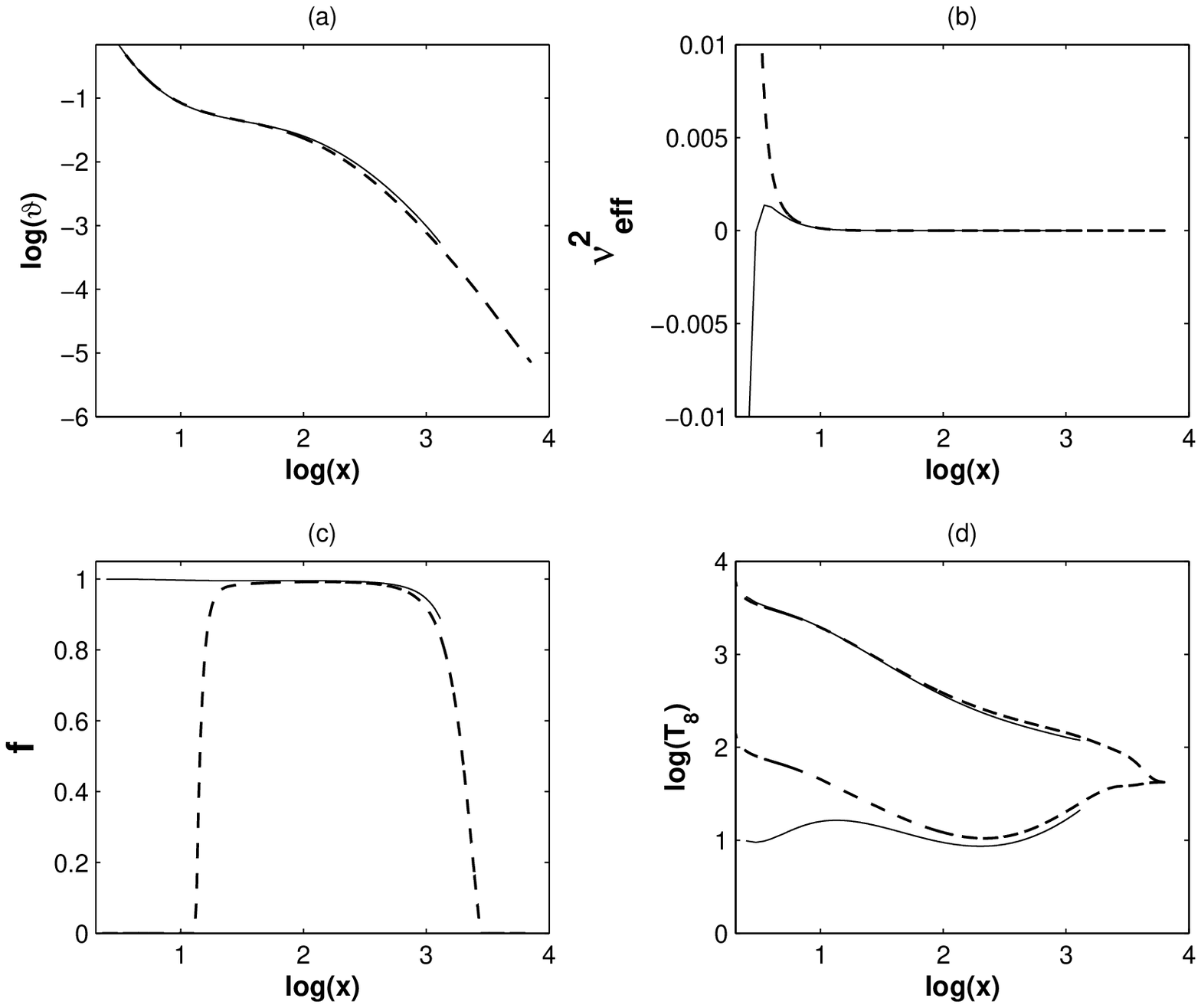}
\caption{
Comparison between solutions for high and low $\alpha$: 
Variation of (a) velocity, (b) square of convective frequency, (c) cooling factor,
(d) ion (upper set of lines) and electron (lower set of lines) temperatures,
a functions of radial coordinate, when solid lines correspond to
$\alpha=0.01$ and dashed lines correspond to $\alpha=0.0001$.
Other parameters are $\mdot=0.01$, $M=10^7$, $a=0$.
}
\label{alfacom}
\end{figure}

Figure \ref{alfacom} shows that although the velocity profiles are similar
for both the values of $\alpha$, the size of the sub-Keplerian disc
is about five times for $\alpha=0.0001$ than that for $\alpha=0.01$.
Inside $x=17$ the low $\alpha$ disc flow becomes cooler very fast, rendering 
$f\rightarrow 0$ at $x>10$ (see Fig. \ref{alfacom}c). 
Therefore, the flow sharply transits from radiatively inefficient in
nature to GAAF. As a consequence,
the low $\alpha$ flow remains stable, as shown in Fig. \ref{alfacom}b, all the way
upto the event horizon. As the sub-Keplerian flow of a smaller $\alpha$
extends further away where the influence of black hole is very weak, 
$T_e$ and $T_i$ merge (see Fig. \ref{alfacom}d) before the flow crosses
the transition radius, unlike the flow with
$\alpha=0.01$ when $T_i>T_e$ there.

\subsection{Comparison between flows around co and counter rotating 
black holes}

As we already discussed that the model with a low mass accretion rate around a supermassive 
black hole is a potential case in explaining the observed dim source Sgr~$A^*$.
On the other hand, the ultra-luminous X-ray sources presumably correspond to 
the models with a high mass accretion flow around a stellar mass black hole. 
Therefore, in order to compare the flow properties between co and counter rotating black 
holes, we choose these two extreme cases.

\begin{figure}
\centering
\includegraphics[width=0.70\columnwidth]{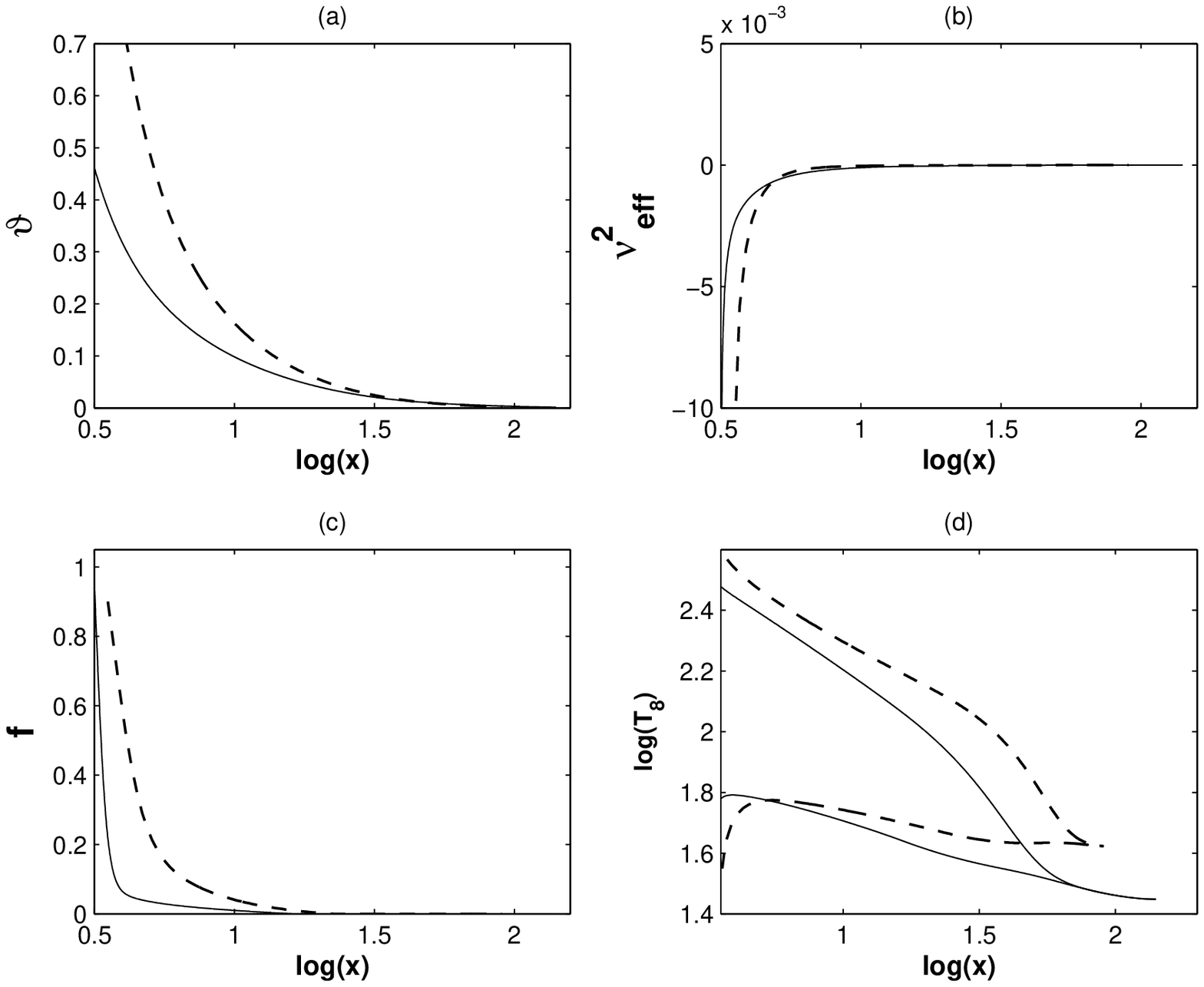}
\caption{
Comparison between solutions for co and counter rotating
stellar mass black holes:
Variation of (a) velocity, (b) square of convective frequency, (c) cooling factor,
(d) ion (upper set of lines) and electron (lower set of lines) temperatures,
as functions of radial coordinate, when solid lines correspond to
$b=0.5$ and dashed lines correspond to $b=-0.5$.
Other parameters are $\mdot=100$, $M=10$, $\alpha=0.01$.
}
\label{kercom1}
\end{figure}

\begin{figure}
\centering
\includegraphics[width=0.70\columnwidth]{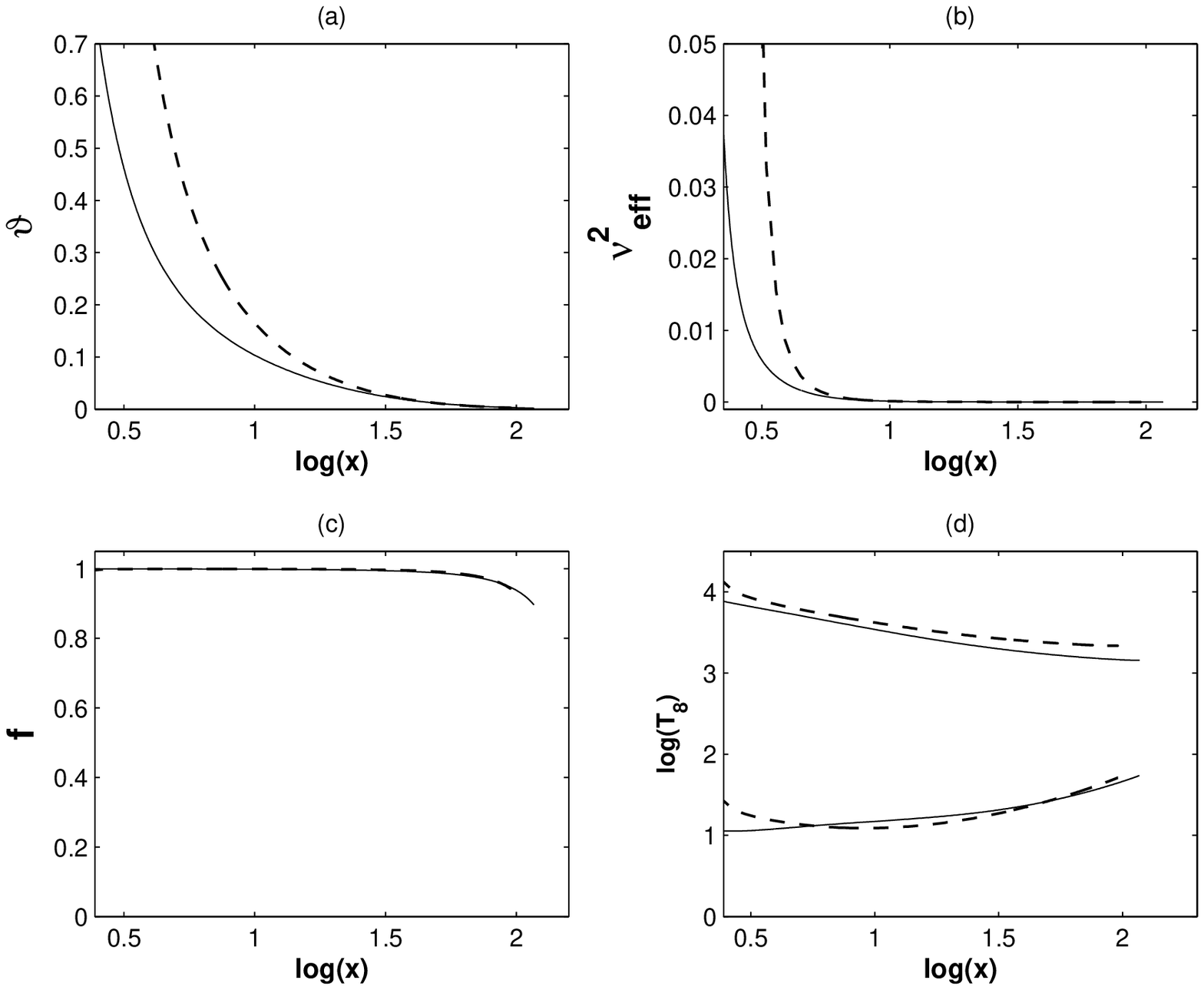}
\caption{
Same as Fig. \ref{kercom1}, except $\mdot=0.01$, $M=10^7$.
}
\label{kercom2}
\end{figure}

Qualitatively, the flows with similar initial conditions around co and counter rotating 
black holes of same mass are similar, as shown in Figs. \ref{kercom1}, 
\ref{kercom2} for $a=\pm0.5$. 
However, the sub-Keplerian disc size around the black hole with $a=-0.5$
is smaller due to smaller value of the effective angular momentum 
(Mukhopadhyay 2003) of the system. Hence the radial velocity is 
almost an order of magnitude higher, particularly at the inner edge, 
for $a=-0.5$. 

\section{Discussion and Summary}

We model the two temperature accretion flow, particularly around
black holes, combining the equations of conservation
and comprehensive cooling processes. We consider self-consistently
the important cooling mechanisms: bremsstrahlung, synchrotron and inverse
Comptonization due to synchrotron photons, where ions and electrons 
are allowed to have different temperatures.
As matter falls in, hot electrons cool
through the various cooling mechanisms, particularly by the synchrotron emission
when the magnetic field is high. This is particularly the case for the
flow around stellar mass black holes where the magnetic field 
may also act as a boost in transporting the angular momentum. 
However, in the present paper, we do not consider such processes
in detail, rather stick with the standard $\alpha$-prescription.

By solving a complete set of disc equations we show that in 
general the disc system exhibits GAAF. However, in certain circumstances
GAAF becomes radiatively inefficient, 
depending on the flow
parameters and hence efficiency of cooling mechanisms. Transitions
from GAAF to radiatively inefficient flow and vice versa are clearly explained by the cooling 
efficiency factor $f$, shown in each model cases. While the previous
authors, who proposed ADAF (Narayan \& Yi 1994, 1995), especially 
restricted with flows having $f=1$ (inefficient cooling), here
we do not impose any restriction to the flow parameters 
to start with and let the parameter $f$ to determine
self-consistently as the system evolves. Therefore, our model
is very general whose special case may be understood as a radiatively
inefficient advection dominated flow.

\begin{figure}
\centering
\includegraphics[width=0.50\columnwidth,angle=-90]{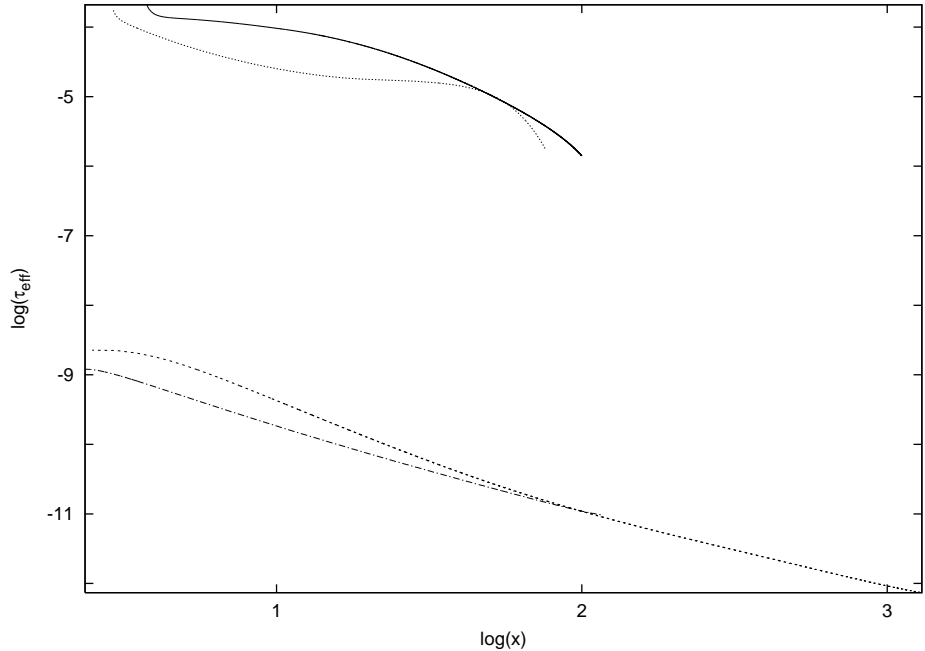}
\caption{
Variation of the effective optical depth as a function of radial coordinate.
Solid ($a=0$) and dotted ($a=0.998$) curves correspond to $M=10$, $\mdot=100$
and dashed ($a=0$) and dot-dashed ($a=0.998$) curves correspond to $M=10^7$, $\mdot=0.01$.
See Tables 1 and 2 for details.
}
\label{opti}
\end{figure}

We have explored especially the optically thin flows incorporating 
bremsstrahlung, synchrotron and inverse Comptonization 
processes. In Fig. \ref{opti} we show the
variation of the effective optical depth as a function of disc radii for two limiting 
cases. While flows around rotating black holes appear slightly thinner compared
to the corresponding cases of static black holes, in general $\tau_{\rm eff}\lsim 5\times 10^{-4}$. 
This verifies our choice of optically thin flows throughout. However, 
for the present purpose, when the main aim is to understand disk dynamics
in the global, viscous, two-temperature regime, we have ignored inverse Comptonization
due to bremssstrahlung photons, if any. This may be important in cases of very super-Eddington
accretion flows which we plan to explore in future, particularly, in analysing the 
underlying spectra.

The temperature of the flow depends on the accretion rate. 
If the accretion rate is low and thus the flow is radiatively 
inefficient, then the disc is hot. Such a hot flow is being
attempted to model since 1976 (Shapiro, Lightman \& Eardley 1976)
when it was assumed that locally $Q^+\sim Q^-$ and thus $f\rightarrow 0$. 
While the model was
successful in explaining observed hard X-rays from Cyg~X-1, it
turned out to be thermally unstable. Rees et al. (1982) 
proposed a hot ion torus model avoiding $f$ to unity. In the
similar spirit Narayan \& Yi (1995) proposed the hot two 
temperature solution in the assumption of $f\rightarrow 1$
including the strong advection into the flow. Abramowicz et al. (1995),
based on the single temperature model, showed that the optically thin disc 
flow of accretion rate more than one Eddington does not have an 
equilibrium solution. However, they did not attempt to solve 
the complete set of differential equations. Based on some simplistic
assumptions they showed the importance of advective cooling. Moreover, 
a single temperature description does not allow them to include all
the underlying physics necessary to describe the cooling processes.
In the due course, Mandal \& Chakrabarti (2005) proposed a two temperature disc
solution where the ion temperature could be as high as $\sim 10^{12}$K.
However, they particularly emphasized on how does the shock in the disc flow
enable cooling through the synchrotron mechanism, without carrying out
a complete analysis of the dynamics. The present paper describes, to our knowledge,
the first comprehensive work to model the two temperature accretion flow self-consistently
by solving the complete set of underlying equations without any pre-assumptive
choice of the flow variables to start with.

The generality lies not only in its construction but also its ability
to explain the under-luminous to ultra-luminous sources, stellar mass
to supermassive black holes. Table 3 lists the luminosities for a
wide range of parameter sets, obtained by our model. It reveals that 
for a very low mass accretion rate $\mdot=0.0001$ around a supermassive
black hole, the luminosity comes out to be $L\sim 10^{34}$ erg/sec, which
indeed is similar to the observed luminosity from a under-luminous 
source Sgr~$A^{*}$. In
other extreme, for $\mdot=100$ around a similar black hole,  
$L\sim 10^{47}$ erg/sec, similar to what observed from the highly luminous AGNs
like PKS~0743-67. On the other hand, when the black hole is considered
to be of stellar mass, then at a high $\mdot=100$, the model reveals
$L\sim 10^{40}$ erg/sec which is similar to the observed luminosity from
ULX sources (e.g. SS433).

\clearpage
\centerline{ Table 3: Luminosity in erg/sec}

\begin{center}
\begin{tabular}{lllllllllllll}
\hline
\hline
$\mdot$ & $M$ &  $\gamma$ & $L$ \\
\hline
\hline
$0.0001$ & $10^7$       & $1.6$ & $10^{34}$ \\
$0.01$ & $10^7$ &  $1.5$ & $10^{38}$  \\
$1$  & $10^7$      & $1.35$ & $5\times 10^{42}$\\
$100$  & $10^7$ &  $1.34$ & $10^{47}$\\
\hline
\hline
$0.01$  & $10$  & $1.5$ & $10^{33}$\\
$1$  & $10$ &  $1.35$ & $7\times 10^{36}$\\
$100$  & $10$   & $1.34$ & $10^{40}$\\
\hline
\hline
\end{tabular}
\end{center}

In general, an increase of accretion rate increases density of the flow
which may lead to a high rate of cooling and thus decrease of the cooling factor $f$.
Hence, $f$ is higher, close to unity reassembling radiatively inefficient flows,
for sub-Eddington accretors,  
and is lower, sometimes close to zero, for super-Eddington flows. Actual value of 
$f$ in a flow also depends on the behaviour of hydrodynamic variables which
determine the rate of cooling processes. Naturally, as the flow advances from 
the transition region to the event horizon, $f$ varies between
$0$ and $1$. However, if the black hole is considered to be rotating,
the flow angular momentum decreases and thus the radial velocity increases.
This in turn reduces the residence time of the sub-Keplerian flow 
hindering cooling processes to complete. This flow is then expected
to be hotter and hence $f$ to be higher compared to that around 
a static black hole. Therefore, the system may tend to be radiatively inefficient, evenif
its counter part around a static black hole appears to be an GAAF.
However, this also depends on the value of $\alpha$, as shown in 
Fig. \ref{alfacom}. A low value of $\alpha$ increases the residence time 
of matter in the disc which helps in cooling processes to
complete, rendering a radiatively inefficient flow to switch over to GAAF. 
This feature may help
in understanding the transient X-ray sources.

In all the cases, the ion and electron temperatures merge or
tend to merge at around transition radius. This is because,
the electrons are in thermal
equilibrium with the ions and thus virial around the transition radius,
particularly when $\mdot\gsim 1$. As the sub-Keplerian flow advances,
the ions become hotter and the corresponding temperature increases, 
rendering the ion-electron Coulomb collisions weaker.
The electrons, on the other hand, cool down via processes like bremsstrahlung,
synchrotron emissions etc. keeping the electron temperature roughly constant
upto very inner disc. This reveals the two temperature flow strictly.

Important point to note is that we have assumed throughout the coupling 
between the ions and electrons is due to the Coulomb scattering. However,
the inclusion of possible nonthermal processes of transferring energy
from the ions and electrons (Phinney 1981, Begelman \& Chiueh 1988)
might modify the results. However, as argued by Narayan \& Yi (1995),
the collective mechanism discussed by Begelman \& Chiueh (1988)
may dominate over the Coulomb coupling at either a very low $\alpha$ or 
a very low $\mdot$. Instead, the viscous heating rate of ions is much larger
than the collective rate of nonthermal heating of electrons, unless
$\alpha$ is too small what we have not considered in the present cases.
Therefore, the assumption to neglect nonthermal heating of
electrons is justified. 

Now the future jobs should be to understand the radiation emitted
by the flows discussed here and to model the corresponding spectra.
This will be the ultimate test of the model in order to explain
observed data.

\appendix
\section{Discussion of boundary values}

We have four coupled nonlinear differential equations (6), (10), (12), (17) 
to be solved for $\vartheta$, $c_{s}$, $\lambda$, $T_{e}$; equations also
involve $\rho$ and $P$. To eliminate $\rho$ and $P$,
we use mass transfer equation (1) and equation of state
(9). Therefore, in total we have five differential equations supplemented 
by an equation of state.
Hence, we need five boundary conditions to start integration. Equation 
(1) can be integrated to obtain
$\mdot$ already given in Eqn. (2), which is supplied as 
an input parameter. Similarly, integrating Eqn. (10) we can obtain 
angular momentum flux 
\begin{eqnarray}
\mdot(\lambda-\lambda_{\rm in})=-4\pi x^2|W_{x\phi}|,
\end{eqnarray}
where $\lambda_{\rm in}$ is the specific angular momentum at the
inner edge of the disc, to be fixed by no torque inner boundary condition. 
Note that $\lambda_{\rm in}\le\lambda_c$ (see, e.g., Chakrabarti 1996). 

We therefore need the initial values of
$\vartheta$, $c_{s}$, $\lambda$ and $T_{e}$ to solve the set of equations. When
we impose the condition that the flow must pass through a critical radius $x_c$
(around a sonic radius) where $D=0$, $\vartheta$ and $c_{s}$ at $x_c$ are related 
by a quadratic equation of Mach number given by Eqn. (24).

For the continuity of $d\vartheta/dx$, $N=0$ at $x_c$.
Therefore, from $N=0$ which an algebraic
nonlinear equation, $c_{s}$ at $x_c$ can be computed iteratively 
(using bisection method), which in turn fixes $\vartheta$ at $x_c$ as well,
provided $\lambda$ is known at that radius.

\begin{figure}
\centering
\includegraphics[width=0.50\columnwidth,angle=-90]{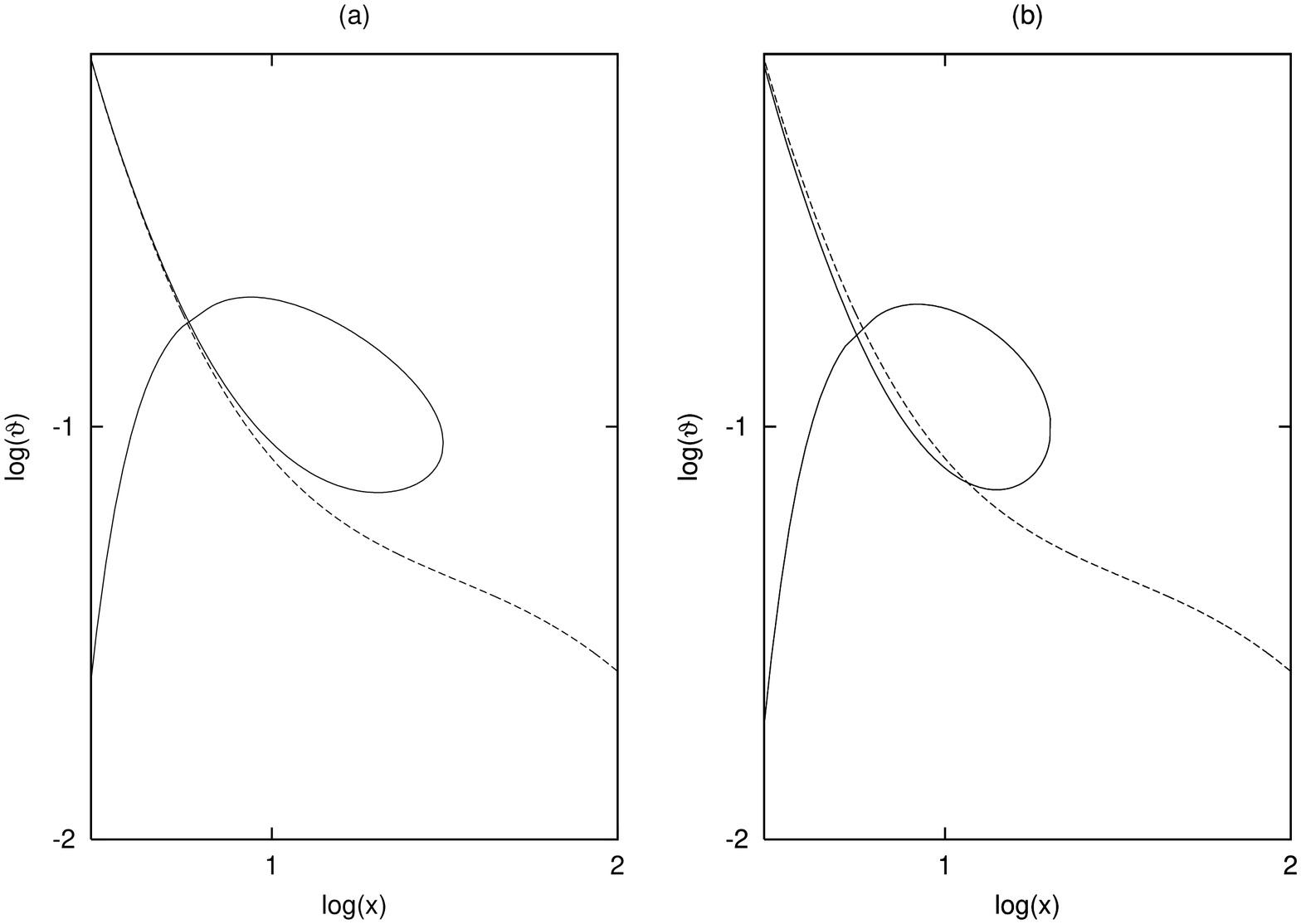}
\caption{
Comparison of the variation of radial velocity as a function of 
radial coordinate (a) between solutions with $x_c=5.5$ (dotted
curve) and $x_c=5.7$ (solid curve), when $\lambda_c=3.2$, 
(b) between solutions with $\lambda_c=3.2$ (dotted curve), $x_c=3.3$ (solid curve), when 
$x_c=5.5$. Other parameters are same as that in Fig. \ref{figstl0}
for $\mdot=0.01$.
}
\label{apen}
\end{figure}

Now we need to set appropriate values of $x_c$ and corresponding 
specific angular momentum $\lambda_c$. This 
is fixed iteratively by invoking the condition that the critical point to be saddle-type. 
This can be seen as follows. First we impose that $\lambda_{\rm in}=\lambda_c$. 
Then by fixing the value of  $\lambda_{c}$ we find
that if $x_c$ is greater than a certain
critical value $x_{cc}$, then the type of $x_c$ changes from saddle-type to $O$-type
which matter never can pass through. Figure \ref{apen}a 
shows how the type of
critical point and then solution topology change with a slight increase of $x_c$.
On the other hand, as $x_c$ decreases from $x_{cc}$ which corresponds
to the energy at $x_c$ ($E_c$) increases,
the sub-Keplerian disc decreases in size. This advances the Keplerian
disc. The reason is that increasing $E_c$ corresponds to decreasing $x_c$
and then increasing centrifugal energy ($\lambda^2/2x|_c$) which keeps the flow Keplerian 
until inner region. However, in principle the solution of the model equations is
possible to obtain for any value of $x_c$ 
from $x_{cc}$ to the marginally bound orbit $x_b$.

Once $x_{c}$ is fixed at $x_{cc}$, we have to obtain the best value of $\lambda_c$.
By increasing the value of $\lambda_{c}$ beyond a certain
critical value $\lambda_{cc}$ at a particular $x_c$, 
we again find a transition from saddle-type to $O$-type critical point. 
Figure \ref{apen}b 
shows how the type of critical point and then solution topology 
change with a slight increase of $\lambda_c$.
On the other hand, decreasing $\lambda_{c}$ from $\lambda_{cc}$
will tend the disc to more Bondi-type. Now for $\lambda_{cc}$ we again
have to obtain a new value of $x_{cc}$ following the procedure outlined
above and thereafter corresponding $\lambda_{cc}$. This needs to be continued 
iteratively until
a specific combination of critical radius $x_{cc}$ and corresponding 
specific angular momentum $\lambda_{cc}$, lying in a narrow range, 
is obtained which leads to a physically interesting large sub-Keplerian 
accretion disc, when matter infalling from a largest possible 
transition radius to a black hole event
horizon through a saddle-type critical point. 
However, in principle the solution of the model equations is
possible to obtain for a range of $\lambda_c$ such that $\lambda_{cc}\ge\lambda_c>0$. 

There is, however, another initial value namely $T_e$
at $x_c$ ($T_{ec}$) to be assigned. 
Choice of $T_{ec}$ depends on the observed nonthermal radiation
which restricts the value of $T_e$ in general. But this restriction can only
provide an order of magnitude of $T_e$. An exact value of $T_{ec}$
should be obtained iteratively from a plausible range of 
$T_{e}$ at $x_c$ so that the values of $x_{cc}$ and $\lambda_{cc}$ obtained following 
the above mentioned procedure converge.

\section{Computation of derivative of velocity at the critical radius}

We first recall the derivative of velocity from Eqn. (20)
\begin{eqnarray}
\frac{d\vartheta}{dx} \,= \,\frac{N(x,\vartheta,c_{s},\lambda,T_{e})}{D(\vartheta,c_{s})}.
\end{eqnarray}
At the critical radius
\begin{eqnarray}
\frac{d\vartheta}{dx} \,= \,\frac{0}{0}.
\end{eqnarray} 
Therefore, we apply l'Hospital's rule and obtain
\begin{eqnarray}
\nonumber
\frac{d\vartheta}{dx} &= &\frac{\frac{D}{Dx}\left[N(x,\vartheta,c_{s},\lambda,T_{e}
)\right]}{\frac{D}{Dx}\left[D(\vartheta,c_{s})\right]}\\
&= &\frac{\frac{dN}{dx} \,+ \,\frac{dN}{d\vartheta}\frac{d\vartheta}{dx} \,+ \,\frac{dN}{dc_{s}}\frac{dc_{s}}{dx} \,+\frac{dN}{d\lambda}\frac{d\lambda}{dx} \,+\frac{dN}{dT_{e}}\frac{dT_{e}}{dx}} 
{\frac{dD}{d\vartheta}\frac{d\vartheta}{dx} \,+ \,\frac{dD}{dc_{s}}\frac{dc_{s}}{dx}}.  
\end{eqnarray}
Now combining with Eqns. (26), (27), (28) we obtain
\begin{eqnarray}
\frac{d\vartheta}{dx} &=&
\frac{N_{1} \,+ \,N_{2} \frac{d\vartheta}{dx}}{D_{1} \,+ \,D_{2}\frac{d\vartheta}{dx}},
\label{dvdxcs}
\end{eqnarray}
where
\begin{eqnarray}
N_{1} \,= \,\frac{dN}{dx} \,+ \,\frac{dN}{dc_{s}} \frac{J}{c_{s}} \,+ \,\frac{dN}{d\lambda}
\left(\frac{c_{s}^{2}-2x\alpha J}{c_{s}}+\vartheta\right) \,+ \,\frac{dN}{dT_{e}}{\frac{(\Gamma_{3}-1)4\pi}{\mdot}\frac{c_{s} x^{3/2}}{F^{1/2}} \left(Q^{ie}-Q^{-}\right)
(1-\Gamma _{1})T_{e}\left(\frac{J}{c_{s}^{2}}+G\right) \,},
\end{eqnarray}
\begin{eqnarray}
N_{2} \,= \,\frac{dN}{d\vartheta} \,+ \frac{dN}{d c_{s}}\left(\frac{c_{s}}{\vartheta}-\frac{\vartheta}{c_{s}}\right) \,+ \frac{dN}{d\lambda}\left(\frac{2 \alpha x}{\vartheta c_{s}} \frac{I_{n+1}}{I_{n}}
\left(\frac{c_{s}^{3}}{\vartheta}-\vartheta c_{s}\right)+\alpha x\right)+ \,\frac{dN}{dT_{e}}(1- \Gamma _{1})T_{e} \frac{\vartheta}{c_{s}^{2}},
\end{eqnarray}
\begin{eqnarray}
D_{1} \,= \,\frac{dD}{d c_{s}} \,\frac{J}{c_{s}}
\end{eqnarray}
\begin{eqnarray}
D_{2} \,= \,\frac{dD}{d\vartheta} \,+ \,\frac{dD}{dc_{s}}\left(\frac{c_{s}}{\vartheta}-\frac{\vartheta}{c_{s}}\right).
\end{eqnarray}
Finally cross-multiplying in Eqn. (\ref{dvdxcs}) we obtain a quadratic equation
\begin{eqnarray}
D_{2}\left(\frac{d\vartheta}{dx}\right)^{2} \,+ \,\left(D_{1}-N_{2}\right)\left(\frac{d\vartheta}{dx}\right)-N_{1} \,= \,0
\end{eqnarray}
such that 
\begin{eqnarray}
\left.\frac{d\vartheta}{dx}\right|_c=\frac{N_2-D_1\pm\sqrt{(D_1-N_2)^2+4D_2N_1}}{2D_2},
\end{eqnarray}
where upper and lower signs correspond to wind and accretion respectively.

\section*{Acknowledgments}
This work is partly supported by a project, Grant No. SR/S2HEP12/2007, funded
by DST, India. One of the authors (SRR) thanks the Council for Scientific and
Industrial Research (CSIR; Government of India) for providing a research fellowship.
The authors also thank the referees for careful reading the manuscript
and providing detailed reports which have helped to improve the paper.

\end{document}